\documentclass[twocolumn,showpacs,preprintnumbers,amsmath,amssymb]{revtex4}

\usepackage{graphicx}
\usepackage{amssymb}
\usepackage{dcolumn}
\usepackage{bm}
\usepackage{subfigure}
\usepackage{amsmath}
\usepackage{color}
\usepackage{ulem}

\begin{document}

\title{Effects of demographic stochasticity on biological community assembly on evolutionary time scales}

\author{Yohsuke Murase}
\author{Takashi Shimada}
\author{Nobuyasu Ito}
\affiliation{
Department of Applied Physics, School of Engineering, The University of Tokyo, 
7-3-1 Hongo, Bunkyo-ku, Tokyo 113-8656, Japan
}

\author{Per Arne Rikvold}
\affiliation{
Center for Materials Research and Technology and Department of Physics, 
Florida State University, Tallahassee, Florida 32306-4350, USA
}

\date{\today}

\begin{abstract}
We study the effects of demographic stochasticity on the long-term dynamics of biological coevolution models of community assembly. 
The noise is induced in order to check the validity of deterministic population dynamics. 
While mutualistic communities show little dependence on the stochastic population fluctuations, 
predator-prey models show strong dependence on the stochasticity, indicating the relevance of the finiteness of the populations.
For a predator-prey model, the noise causes drastic decreases in diversity and total population size. 
The communities that emerge under influence of the noise consist of species strongly coupled with each other 
and have stronger linear stability around the fixed-point populations than the corresponding noiseless model.
The dynamics on evolutionary time scales for the predator-prey model are also altered by the noise. 
Approximate $1/f$ fluctuations are observed with noise, while $1/f^2$ fluctuations are found for the model without demographic noise.
\end{abstract}

\pacs{87.23.Kg,05.40.-a,05.65.+b}


\maketitle



\section{Introduction}\label{sec:introduction}

Noise may be a relevant perturbation to many kinds of population dynamics.
Effects of population fluctuations have been investigated for several models, 
for example, predator-prey models \cite{mckane2005ppc,Pineda-Krch:2007eu}, epidemic models \cite{alonso2007sae}, the Ricker model \cite{Melbourne:2008nx},
evolutionary game theories \cite{reichenbach:051907,Reichenbach:2007wd, PhysRevLett.95.238701,traulsen:011901,Nowak:2004vf,Taylor20041621}, 
and pattern formation \cite{mobilia:040903,Mobilia:2007nx,Butler:2009sy}. 
Since the birth-death process of individuals is stochastic, 
the population of each species always fluctuates due to the finiteness of the number of individuals, 
known as demographic stochasticity. 
Demographic stochasticity is an endogenous phenomenon, 
and the species populations may fluctuate, even in a constant environment. 
Population dynamics with demographic stochasticity are more realistic 
than the corresponding deterministic description which is valid for infinite populations in constant environments, 
and they often show nontrivial dynamics which cannot be predicted by the deterministic equations. 
For a particular predator-prey system \cite{mckane2005ppc}, 
the demographic stochasticity causes oscillations, 
while the system is asymptotically stable under the corresponding deterministic population dynamics. 
Since the demographic stochasticity effectively adds uncorrelated noise to the population dynamics, 
the oscillations at the eigenfrequencies are amplified by a large factor. 
Hence, the effect of the demographic stochasticity is much larger 
than the one estimated by naive $O(1/\sqrt{N})$ estimates. 
For a neutrally stable system \cite{reichenbach:051907}, the noise effect becomes even more drastic: 
only one species can survive and the others die out after a sufficiently long time, 
while the corresponding deterministic model predicts the coexistence of the species, with regular oscillations. 
Stochasticity may also influence the outcome of the evolutionary dynamics.
In small populations, the evolutionary branching is delayed compared to the case of larger populations, 
and the delay strongly depends on the absolute population size \cite{claessen2007delayed}.
Several empirical data sets are also compared with theoretical models and 
are described better by models with stochastic population dynamics \cite{Melbourne:2008nx,Pineda-Krch:2007eu}.
Thus, population fluctuations, which inevitably exist in any finite system, may 
drastically alter the predictions of deterministic models 
and often cause decreases in biodiversity.
A major goal of this paper is to investigate the effects of demographic stochasticity 
in models of biological community assembly on evolutionary time scales. 

Several models to bridge ecological and evolutionary time scales have been suggested, 
such as the tangled-nature model \cite{PhysRevE.66.011904,CHRISTENSEN:2002yq,0305-4470-36-4-302}, 
simplified versions of that model \cite{rikvold2003pea,zia-jpa,0305-4470-38-43-005,Rikvold:2007lr,rikvold2007ibp}, 
the Webworld model \cite{Caldarelli:1998eu,drossel01:_influen_of_predat_prey_popul,Drossel:2004fj,McKane:2004qy}, 
the scale-invariant model \cite{shimada-arob2002}, and others \cite{PhysRevLett.90.068101,tokita-tpb2003}. 
More concretely speaking, 
these are population dynamics models with additional rules for the introduction and extinction of species. 
New species, whose interaction coefficients are assigned by a rule, are added to the community at a certain rate; 
and the extinction of resident species can happen due to the population dynamics. 
Evolution is modeled by repeating the introductions and the extinctions of species in these models. 
Potential numbers of species are much larger than the number of species coexisting at the same time.
If the population dynamics have a noise term, the emergent communities can be nontrivially different from 
the communities selected without this noise.
This is because the speciation events that trigger large changes at the population level invariably involve 
single or very small numbers of individuals that are highly susceptible to statistical fluctuations \cite{claessen2007delayed}. 
The main issues we address here are the noise effects on 
(i) the properties of the emergent communities 
and (ii) the statistics of the evolution dynamics, especially the intermittency during evolution.

In this paper, we use the simplified versions of the tangled-nature models and study the effects of demographic stochasticity.
For these models, the fixed point and the linear stability around it for a given community are analytically obtainable. 
This helps us estimate the relevance of the noise in the population dynamics. 
Furthermore, long-term dynamics on evolutionary time scales are studied extensively, 
and $1/f$ fluctuations and power-law duration distributions are found for these models. 
We show that the dynamics observed for the individual-based models 
may undergo qualitative changes from the corresponding models with deterministic population dynamics.

The organization of this paper is as follows. 
In section \ref{sec:models}, the models are defined, 
and some topics related to these models are discussed. 
Results of the simulations are shown in section \ref{sec:results}. 
In section \ref{subsec:diversity-decline}, we show how population fluctuations affect the diversity, 
and in section \ref{subsec:long-term}, we explore the dynamics on evolutionary time scales.
Section \ref{sec:summary} is devoted to a summary and discussion.
Some mathematical details are discussed in Appendices \ref{sec:f-fluc}-\ref{sec:b-b-const}.

\section{Models}\label{sec:models}

The models considered here are extensions of 
the tangled-nature model studied in \cite{rikvold2003pea,rikvold2007ibp,Rikvold:2007lr}.
The tangled-nature model is an individual-based model, originally introduced 
by Hall and co-workers \cite{PhysRevE.66.011904} 
and later simplified by Rikvold and Zia \cite{rikvold2003pea}.
In the simplified models \cite{rikvold2003pea,zia-jpa,0305-4470-38-43-005,rikvold-2005,rikvold2007ibp,Rikvold:2007lr}, 
the population evolves stochastically in discrete, non-overlapping generations. 
In these models, each individual of species $I$ gives rise to $F$ offspring 
with a reproduction probability $P_I$ before it dies. 
Otherwise it dies without offspring. 
The reproduction probability $P_I$ for an individual of species $I$ in generation $t$ 
depends on the individual's ability to utilize the amount $R$ of available external resources, 
and on its interactions with the population sizes $n_J(t)$ of all the species present 
in the community at that time.
The form of $P_I$ is discussed in the following subsection.

In the individual-based models, species populations evolve stochastically.
The probability $p_{I}(k)$ that $k$ out of $n$ individuals of species $I$ succeed in producing offspring 
is given by the binomial distribution, 
\begin{equation}
	p_{I}(k) = \binom{n}{k} P_{I}^{k} (1-P_{I})^{n-k},
\end{equation}
where $\binom{n}{k}$ is the binomial coefficient and $P_{I}$ is the probability that an individual of species $I$ gives rise to offspring in that generation.
Thus, the mean and variance of the number of offspring are $nP_{I}$ and $nP_{I}(1-P_{I})$, respectively.
In this paper, we consider an approximation to the binomial distribution 
by the Gaussian distribution with mean $nP_{I}$ and variance $nP_{I}(1-P_{I})$ in order to control the strength of the stochasticity. 
The following stochastic difference equation is used for the population updates:
\begin{equation}\label{eq:population-update}
	n_I(t+1) = F[ P_{I}n_{I}(t) + \kappa\sqrt{n_I(t)P_I(1-P_I)} \xi(t) ] \;, 
\end{equation}
where $\kappa$ and $\xi(t)$ are a control parameter for the noise strength 
and a Gaussian random number with mean $0$ and variance $1$, respectively. 
When $\kappa=1$, this update algorithm is a good approximation 
for the corresponding individual-based model, 
while it is deterministic when $\kappa=0$.
Although there is no easy interpretation except for $\kappa=0$ and $1$,
we use several intermediate values of $\kappa$ in order to investigate the crossover between deterministic and individual-based models.
The population size $n_I(t)$ is a positive real number 
while it is a positive integer in the original individual-based models. 
The approximation by the Gaussian distribution to the binomial one is known to be quite good 
when $n_I$ is sufficiently large.
Typically when $nP_{I}$ and $n(1-P_{I})$ are greater than five, the approximation is good.
Even when $n_I$ is small, we expect that 
the approximate model still captures the essence of the population fluctuations in the individual-based model, 
although the population dynamics for species with very small populations 
are critical for the emergence or extinction of species. 
Since it is not necessary to draw random numbers for every individual, 
this update rule is computationally more efficient than the true individual-based model 
and enables us to run simulations for longer times.

It is straightforward to extend the model so that the number $F$ of offspring per individual follows a stochastic process.
(See Appendix \ref{sec:f-fluc}.)
In that case, the fluctuations are even more enhanced and the difference from the deterministic models are more important.
In this paper, however, we limit ourselves to the case that $F$ is fixed for simplicity.
Even with this model, the differences between stochastic and deterministic population dynamics are observed as shown later.

To mimic an evolutionary process, 
speciation and extinction of species are introduced.
New species are added to the system by ``mutation'' of resident species.
These rules are formulated in section \ref{subsec:addition}. 
Extinction of a species happens when its population becomes less than a threshold value, $n_{\rm thr} = 0.5$. 
When the $I$th species goes extinct, this species is eliminated from the system, 
i.e., the number of degrees of freedom decreases by one. 
The community configuration reorganizes as a result of the appearance and extinction of species. 

\subsection{Reproduction probability}\label{subsec:tn}

As in the original models \cite{rikvold2003pea}, the reproduction probability $P_I$ is taken as 
\begin{equation}\label{eq:p}
P_I(R,\{n_J(t)\}) = \frac{1}{1+\exp{ \left[ - \Delta_I(R,\{n_J(t)\}) \right] } } \;,
\end{equation}
where
\begin{equation}\label{eq:delta}
\Delta_I(R,\{n_J(t)\}) = - b_I + \frac{\eta_I R}{N_{\rm tot}(t)} 
+ \sum_J \frac{M_{IJ} n_J(t)}{N_{\rm tot}(t)} - \frac{N_{\rm tot}}{N_{\rm 0}}.
\end{equation}
Here $b_I$ is the cost of reproduction for species $I$ (always positive), 
and $\eta_I$ is the ability of individuals of species $I$ to utilize the external resource $R$. 
The interaction matrix $\mathbf{M}$ defines the interactions between species.
The total population size is denoted by $N_{\rm tot}(t) = \sum_J n_J(t)$, 
and $N_0$ is an environmental carrying capacity that prevents $N_{\rm tot}(t)$ from diverging to infinity.
The reproduction probability $P_I(R,\{n_J(t)\})$ is a monotonically increasing function of $\Delta_I$, 
ranging over $(0,1)$.
Thus $\Delta_I$ is a measure of the fitness of species $I$.
For a large positive $\Delta_I$ 
(small birth cost, strong coupling to the external resource, and more prey than predators), 
$P_I$ goes to one and the population of species $I$ increases.
In the opposite limit of large negative $\Delta_I$ 
(large birth cost, weak or no coupling to the external resources, and/or more predators than prey),
$P_I$ goes to zero and the population decreases rapidly. 
The nonlinear dependence of $P_I$ on $\Delta_I$ thus limits the growth rate of the population size, 
even under extremely favorable conditions for species $I$.

Two types of reproduction probabilities are considered in this paper: Model A and Model B. 
Model A has no restriction on the form of the interaction matrix $\mathbf{M}$. 
Therefore, each species makes various types of interactions with others, 
including predator-prey, mutualistic, and competitive interactions.
In contrast, the interspecies interactions are limited to predator-prey interactions in Model B.
This is realized by the limitation that the off-diagonal part of $\mathbf{M}$ 
must be antisymmetric ($M_{IJ} = -M_{JI}$). 
Thus, if $M_{IJ} > 0$ and $M_{JI} < 0$, 
then species $I$ is the predator (or parasite) and $J$ the prey (or host), and vice versa. 
Model A has a more general form, while Model B focuses on the energy transport via the foodweb.

Model A was introduced and studied in \cite{rikvold2003pea,zia-jpa}.
In this model, the reproduction cost $b_I$ and the external resource $R$ are zero; 
thus the first and the second terms of Eq.~(\ref{eq:delta}) disappear: 
\begin{equation}
\Delta_I(\{n_J(t)\}) = \sum_J \frac{M_{IJ} n_J(t)}{N_{\rm tot}(t)} - \frac{N_{\rm tot}}{N_{\rm 0}}.
\end{equation}
The total population size is limited by the last term, which includes the carrying capacity $N_{\rm 0}$.
The off-diagonal elements of the interaction matrix $M_{IJ}$ are randomly drawn 
from a uniform distribution over $[-1,+1]$, 
while the diagonal elements are set to zero.
For Model A, $F=4$ and $N_0 = 2000$ are used in this paper.
The value $F=4$ for Model A is chosen such that the population dynamics for a single 
species should relax monotonically to a stable fixed point 
in order to ensure that any complex dynamics are due to interspecies interactions \cite{rikvold2003pea}.
As shown in \cite{rikvold2003pea,zia-jpa,filotas-TaNaA,Filotas:2010th}, communities tend to evolve toward mutualism in Model A.

In Model B \cite{rikvold2007ibp}, the external resource $R$ is introduced.
All the species have positive values of the birth cost $b_I$, which are randomly drawn from $[0,1]$, 
and a certain proportion (0.05 is used in this paper) of species can feed on the resource, 
i.e., the resource couplings $\eta_I$ are positive for primary producers or autotrophs, 
and zero for consumers or heterotrophs. 
Here an abiotic resource $R$ is introduced that is renewed each generation at the same level
(here, $2000$) and does not have independent dynamics.
The off-diagonal part of the interaction matrix is limited to be antisymmetric.
Non-zero elements are assigned randomly to the pairs of ($M_{IJ}$, $M_{JI}$) 
with probability $c = 0.1$, which is consistent with food webs in nature, such as St. Marks Seagrass, St. Martin Island, and Little Rock Lake \cite{Dunne:2002eu,Dunne:2002oq}.
The nonzero elements of the interaction matrix are randomly chosen 
from a triangular distribution on $[-1,+1]$. 
These parameter ranges were chosen 
to compare with the corresponding individual-based model \cite{rikvold2007ibp}.
The diagonal elements of $\mathbf{M}$, which represent the intraspecies interactions, 
are selected randomly from a uniform distribution on $[-1,0]$ for all the species.
The environmental carrying capacity term is not included in this model ($N_0 = \infty$). 
Thus, 
\begin{equation}
  \label{eq:delta-modelB}
\Delta_I(R,\{n_J(t)\}) = - b_I + \frac{\eta_I R}{N_{\rm tot}(t)} + \sum_J \frac{M_{IJ} n_J(t)}{N_{\rm tot}(t)}.
\end{equation}
The birth cost term and the negative diagonal elements $M_{II}$ 
prevent species populations from growing to infinity. 
The fecundity $F$ for Model B is set to $2$. 
With this value, the dynamics for a single species approaches its fixed point monotonically.
	
These models have fixed points $|n^{\ast}\rangle$, which can be calculated exactly \cite{Rikvold:2007lr}.
Here $|n^{\ast}\rangle$ is a column vector of the equilibrium population sizes of all species present in the community.
Linear stability around this fixed point can also be estimated. 
See Appendix \ref{sec:fp-calculation} for these solutions.

\subsection{Introduction of new species}\label{subsec:addition}
Communities are assembled through mutations of resident species as follows. 
Each species has a bit-string genome of length $L$, thus the total number of potential species is $2^L$. 
All the species-specific values, $b_I$, $\eta_I$ and $M_{IJ}$, 
are predetermined at the beginning of the simulation and fixed during the simulation.
In every generation, a mutation happens to the genomes of the existing offspring at the moment: 
all the bits existing in the system, which amount to $N_{\rm tot}^{\rm ind} \times L$, 
flip independently with a probability $\mu/L$, resulting in the appearance of new species. 
The genomic mutation rate, $\mu$, determines how frequently individuals mutate.
Here, the number of individuals belonging to species $I$, $n^{\rm ind}_{I}$ is calculated by 
rounding off the population size, $n^{\rm ind}_{I} = \lfloor n_I + 0.5 \rfloor$, 
and the total number of individuals is $N_{\rm tot}^{\rm ind} = \sum_I n^{\rm ind}_{I}$.
Thus, an individual moves to a neighbor in the $L$-dimensional hyper-cubic genome space by a mutation. 
The probability of $m$-bit mutations in a single individual is small, $O(\mu^m)$, 
therefore the probability of multi-gene mutation is small.
The coefficients of species $I$ ($b_I$, $\eta_I$, and $M_{IJ}$) 
have no correlation with those of its neighbor species, which is a less realistic aspect of the model.
However, the model captures the aspect that 
the number of mutant species accessible from a given community is limited \cite{murase:2009}. 
Models to overcome this problem have also been proposed, 
and it is confirmed that the phenotypic correlation does not qualitatively alter the long-term fluctuations \cite{0305-4470-38-43-005}.

Each simulation run was started with a single, randomly chosen species
(producer species for Model B) with a population size of $100$ individuals.
The details of this initial condition are totally insignificant,
and the systems were completely ``thermalized''
during the initial warm-up periods.

We also note that the results shown below do not show qualitative dependence 
on the precise values of $L$ and $\mu$ for reasonable ranges.
If the mutation rate is too high, the system shows mutational meltdown and the number of species diverges.
For too short $L$ (typically $L < 10$), the system is trapped in a certain state and the species composition never changes.
We choose parameters so that these unrealistic cases are excluded and simulations are computationally feasible.

\section{Results}\label{sec:results}

In this section, we focus on the effects of the demographic noise on the community structure. 

\subsection{Robustness against the noise}\label{subsec:diversity-decline}
\subsubsection{Model A}\label{subsubsec:diversity-decline-modelA}
\begin{figure}[ht!]
\begin{center}
\subfigure{
\includegraphics[width=8.0cm]{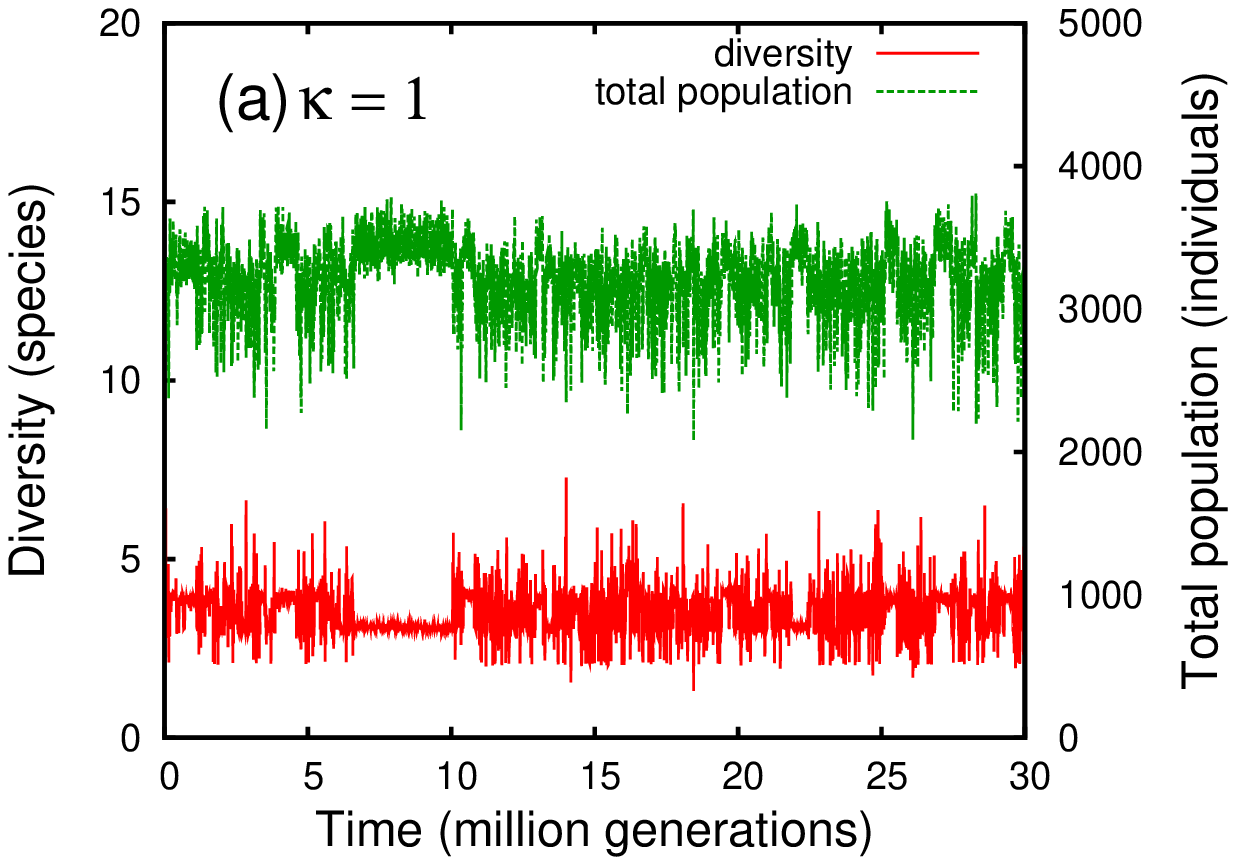}
\label{fig:a-mut-10-ts}
}
\subfigure{
\includegraphics[width=8.0cm]{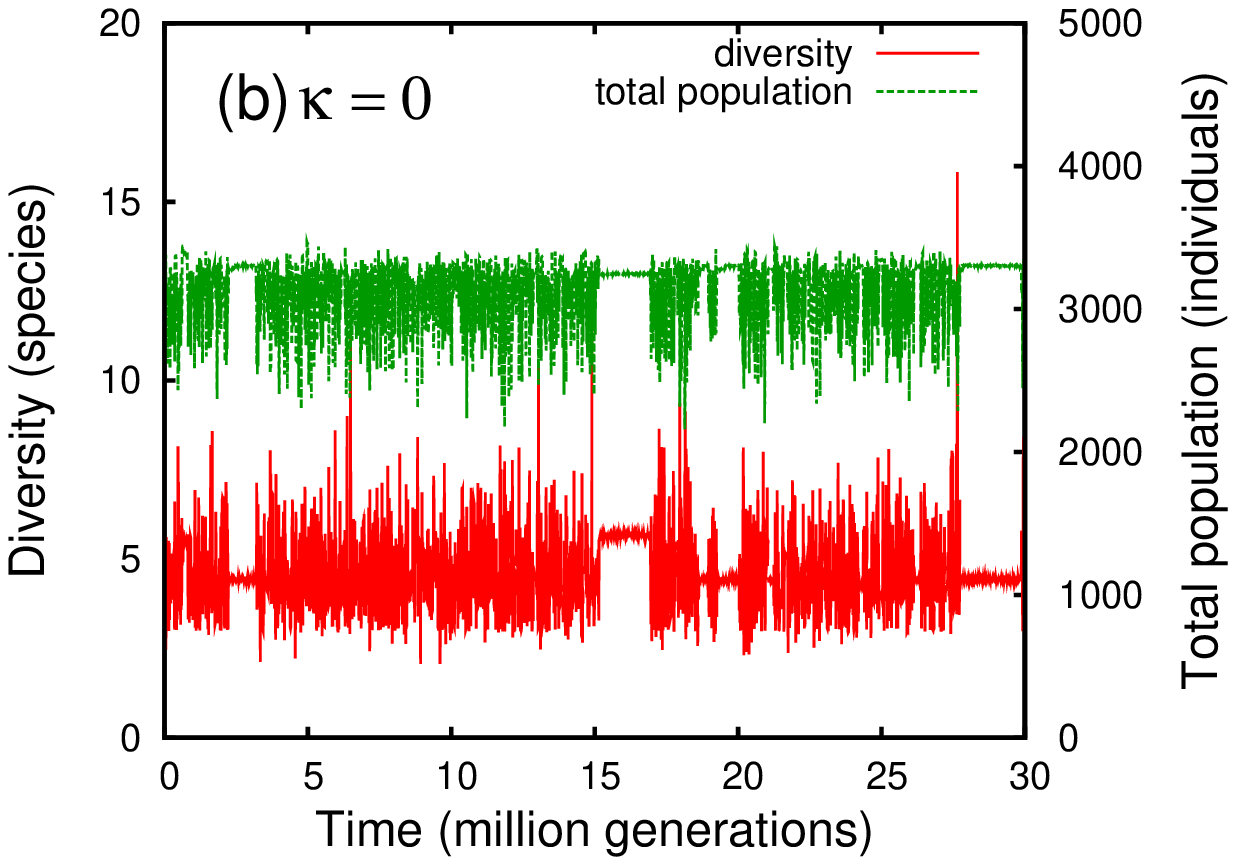}
\label{fig:a-mut-00-ts}
}
\end{center}
\caption{
(Color online) Typical time series of the exponential Shannon-Wiener diversity and the total population size 
plotted every $8000$ generations 
for Model A with $\mu = 0.001$, $L = 13$, and (a) $\kappa=1$ and (b) $\kappa=0$.
The upper and lower curves in the figures show total population size and diversity, respectively.
\label{fig:a-mut-timeseries}}
\end{figure}

First we show how the population fluctuations affect the growth of the community diversity.
Figure \ref{fig:a-mut-timeseries} shows the time evolution of the diversity index $D$ 
and the total population size $N_{\rm tot}$
for Model A with $\kappa = 1$ and $0$, respectively.
Here, the diversity index is defined as $D = \exp{(S)}$, where
\begin{equation}\label{eq:diversity}
	S\left( \left\{ n_I\left(t\right) \right\} \right) 
	= -\sum_{ \left\{I|\rho_I(t)>0\right\} } \rho_I(t)\ln{ \rho_I(t) } 
\end{equation}
with $\rho_I(t) = n_I(t)/N_{\rm tot}(t)$, is the information-theoretical entropy of the species distribution. 
The quantity $D$ is known as the exponential Shannon-Wiener diversity index \cite{Krebs:1989pi}.
This measure can be interpreted as an effective number of species, and so it has the same units (species) as the species richness.
We adopt this measure in order to filter out unsuccessful mutants which have tiny populations and rapidly go extinct.
It has been confirmed that $D$ is approximately proportional to the number of species constituting ``core communities,''
composed of species who succeeded to have a positive stable fixed-population \cite{rikvold2007ibp}.
Here we do not use a finite-size correction \cite{Schurmann:2004kx} for the estimation of $S$ for simplicity.
This correction is numerically confirmed to be less than one percent and therefore we here adopt the simpler form [Eq.~(\ref{eq:diversity})].

The mutation rate $\mu = 0.001$ and the genome length $L=13$ are used for Model A. 
The difference of the average $D$ and $N_{\rm tot}$ between $\kappa=0$ and $1$ is slight. 
Both figures show similar intermittent behaviors, 
consisting of active and quiet periods. 
During the active periods, the diversity measure and the total population size show larger fluctuations, 
and the species composition changes quickly. 
On the other hand, during the quiet periods, the species composition remains nearly constant, 
and the system is considered to be in a quasi-steady state (QSS). 
The evolution proceeds intermittently, rather than gradually, like a stick-slip motion, repeating active and quiet periods. 
Average diversity and total population size for Model A are summarized in Table \ref{tab:mut_A_average}. 
Both measures decrease moderately with increasing noise level. 

\begin{table}[ht!]
\begin{center}
\caption{
Numerical results for Model A. The data are averaged over twelve independent runs.
The initial $2^{24}$ generations are considered as a ``warm-up'' period and not included in the statistics. 
The statistical errors are shown in parentheses. 
In this paper, the statistical error of the value $x$ is calculated as $\sqrt{ \left(\sum{(x-\langle x\rangle)^2}\right) / (n(n-1)) }$, where $n$ is the number of independent runs. 
}
\label{tab:mut_A_average}
\begin{tabular}{c|c|c}
\hline\hline
$\kappa$ & $\overline{D}$ (species) & $\overline{N_{\rm tot}}$ (individuals) \\\hline
1 & 3.55 (3) & 3155 (11) \\
0.1 & 4.81 (11) & 3246 (23) \\
0 & 4.55 (9) & 3337 (36) \\\hline\hline
\end{tabular}
\end{center}
\end{table}

\subsubsection{Model B}\label{subsubsec:diversity-decline-modelB}
\begin{figure}[ht!]
\begin{center}
\subfigure{
\includegraphics[width=8.0cm]{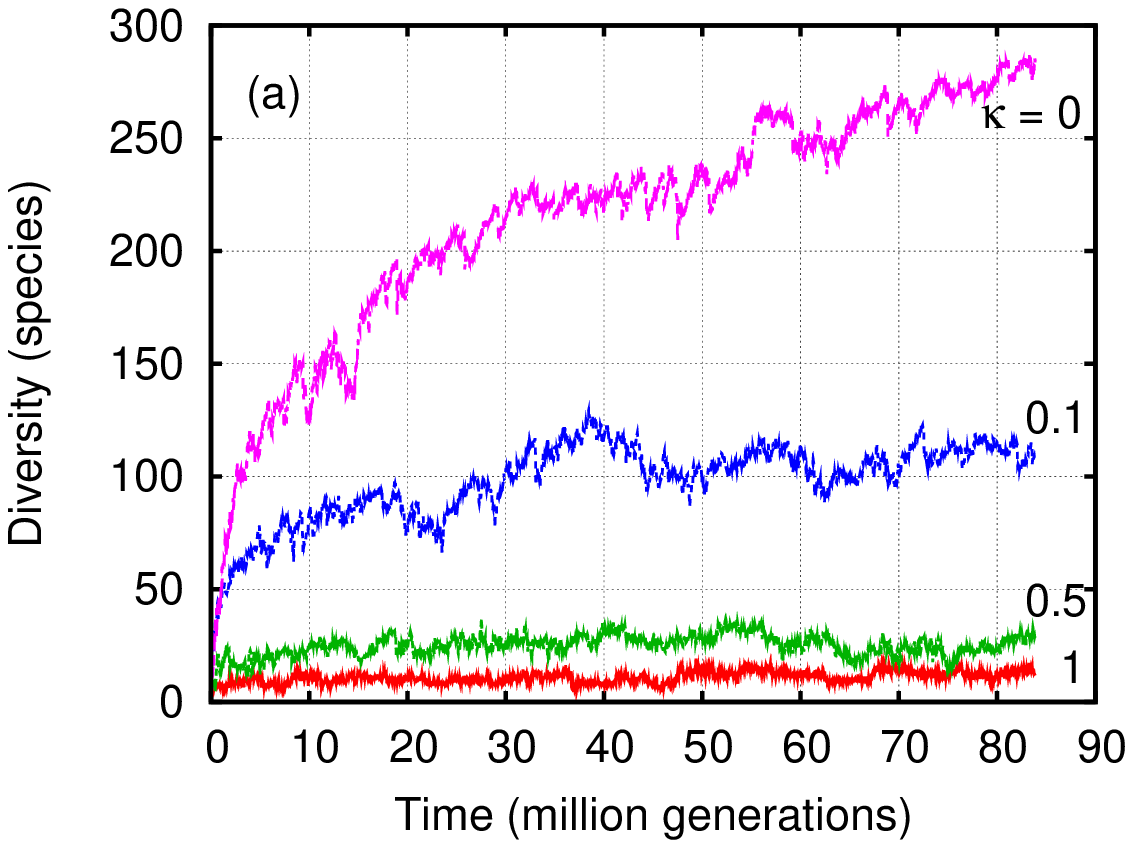}
\label{fig:b-mut-div}
}
\subfigure{
\includegraphics[width=8.0cm]{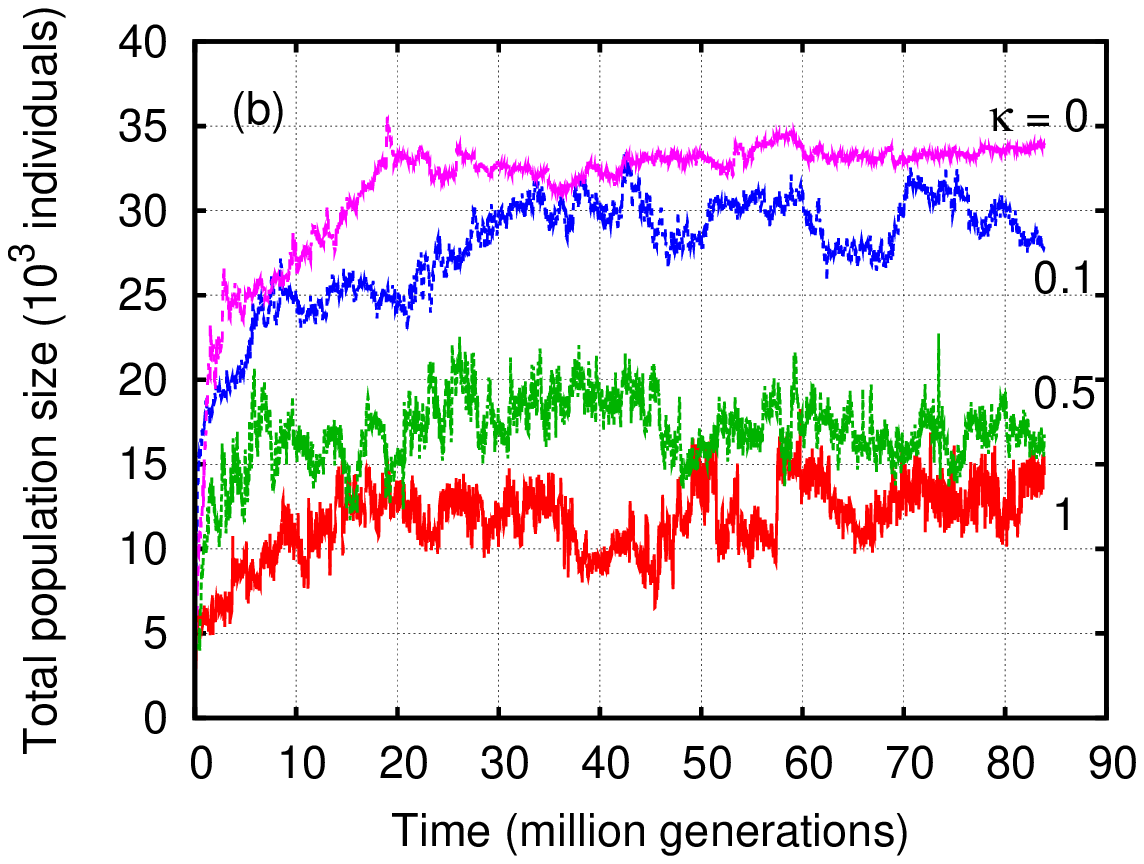}
\label{fig:b-mut-biomass}
}
\end{center}
\caption{
(Color online) Time series of (a) exponential Shannon-Wiener diversity and (b) total  population size 
for Model B with $\mu = 0.0005$ at several noise levels. 
The data are plotted every $8192$ generations.
In either figure, 
the curves correspond to $\kappa=0$, $0.1$, $0.5$, and $1.0$ from top to bottom, respectively.
\label{fig:b-mut-div-biomass}}
\end{figure}

On the other hand, for Model B, the dependence on $\kappa$ is remarkable.
Figure \ref{fig:b-mut-div-biomass} shows typical time series of the diversity index 
and the total population size for Model B at several noise levels with $\mu = 0.0005$ and $L=18$. 
The data are plotted every $8192$ generations for improved visibility. 
As the noise level $\kappa$ increases, 
both the diversity and the total population size decrease remarkably.
For other $\mu$ and $L$, strong dependence on the noise is also observed for Model B.
Thus, Model A and Model B show fundamental differences.
This means that it gets more difficult for species to survive, due to the stochastic population fluctuations. 
Although the species populations basically fluctuate around their fixed points, 
the probability that a population size touches the extinction threshold increases 
under strong stochastic population fluctuations. 
When $\kappa$ is small, high diversity and high total population size are realized. 
Especially, when $\kappa=0$, the system does not reach a statistically stationary state, 
even after $80$ million generations. 
In addition, the fluctuations are less intermittent than for $\kappa=1$.
This intermittency will be quantitatively estimated in the next subsection.
We also note that the diversity for $\kappa=1$ is approximately the same as that of the individual-based model \cite{rikvold2007ibp}, 
indicating that the current model with $\kappa=1$ is a good approximation to the individual-based one. 
The averages of the diversity and total population size for several values of $\kappa$ are summarized in Table \ref{tab:mut_B_average}.

\begin{table}[htbp]
\begin{center}
\caption{
Numerical results for Model B. The data are averaged over six independent runs. 
The initial $2^{24}$ generations are considered as a ``warm-up'' period and not included in the statistics. 
The statistical errors are shown in parentheses. 
Uncertainties of $\overline{N_{\rm tot}}$ are also in units of $100$ individuals. 
}
\label{tab:mut_B_average}
\begin{tabular}{c|c|c}
\hline\hline
$\kappa$ & $\overline{D}$ (species) & $\overline{N_{\rm tot}}$ ($100$ individuals) \\\hline
1 & 12.2 (6) & 127 (8) \\
0.5 & 24.7 (4) & 164 (4) \\
0.1 & 101 (2) & 278 (5) \\
0 & $-$ & 315 (6) \\
\hline\hline
\end{tabular}
\end{center}
\end{table}

\begin{figure}[ht!]
\begin{center}
\subfigure{
\includegraphics[width=8.0cm]{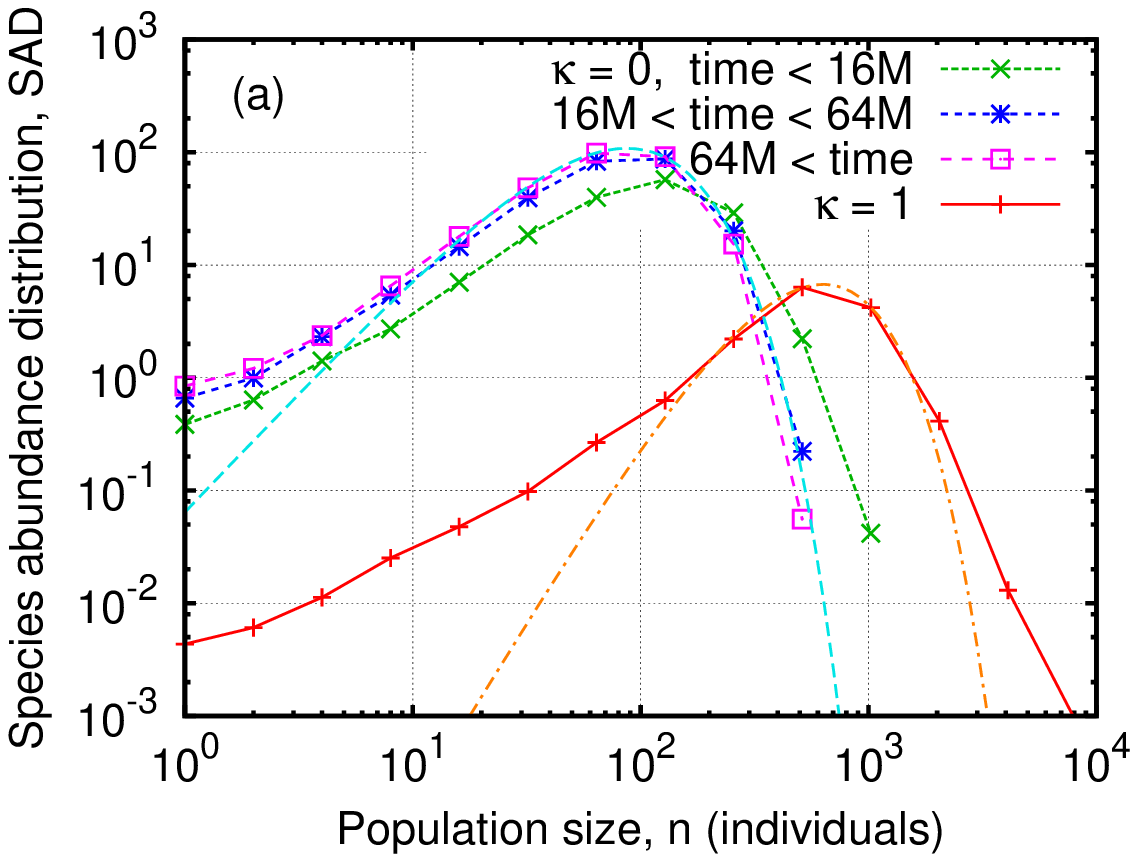}
\label{fig:b-sad}
}
\subfigure{
\includegraphics[width=8.0cm]{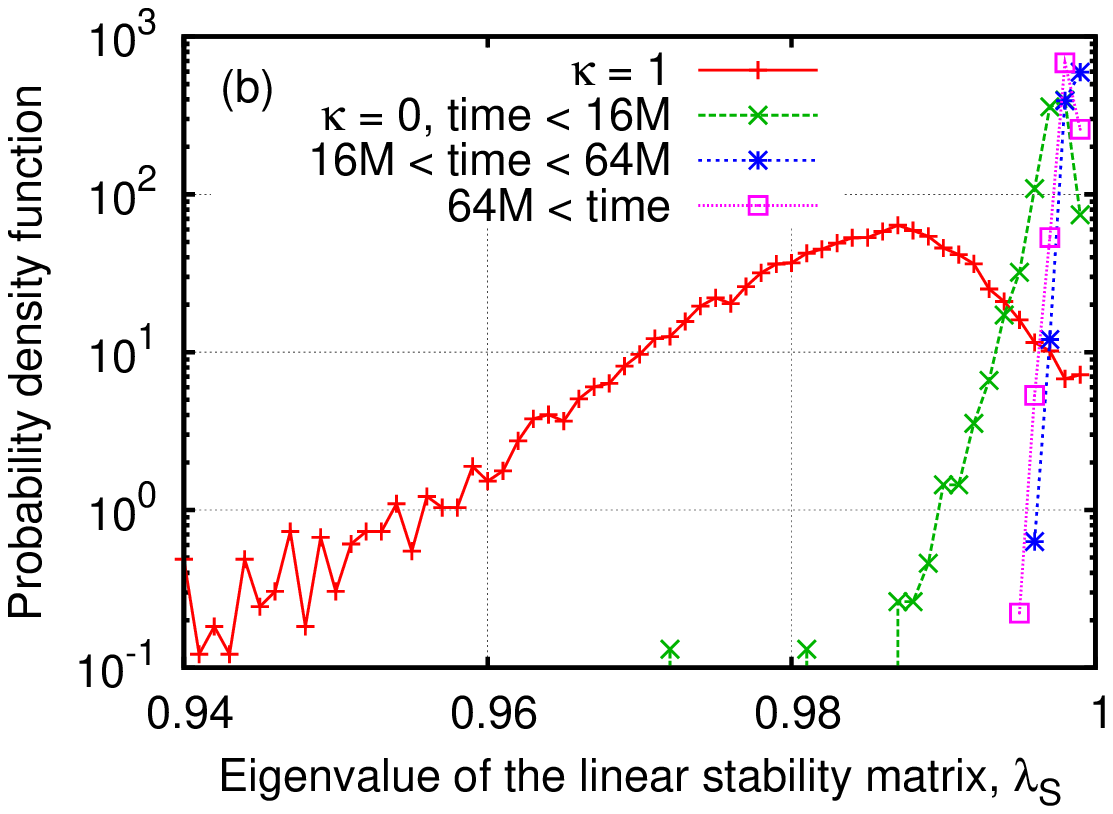}
\label{fig:b-s-eigens}
}
\subfigure{
\includegraphics[width=8.0cm]{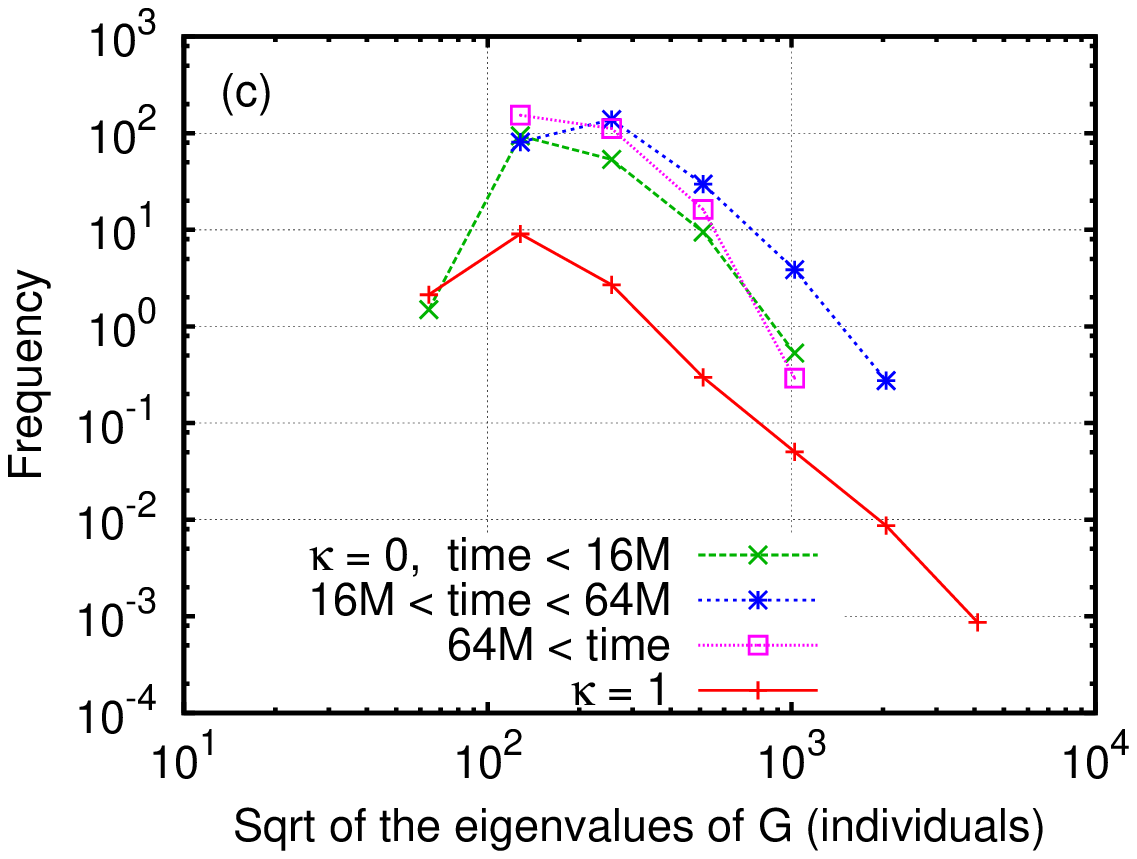}
\label{fig:b-g-eigens}
}
\end{center}
\caption{
(Color online) 
(a) Species abundance distribution (SAD), 
(b) probability density functions (pdf) of the eigenvalues of the linear stability matrix, $\lambda_{\mathbf{S}}$,  
(c) frequency of the eigenvalues of the covariance matrix, $\lambda_{\mathbf{G}}$,  
 for Model B with $\kappa = 1$ and $\kappa = 0$. The data are normalized in the same way as the SADs.
For $\kappa = 0$, the data are obtained for three time intervals. 
The SADs show the number of species whose populations are within each bin region, for each community. 
The data were sampled every one million generations, and averaged over $18$ independent runs. 
The fitting functions Eq.~(\ref{eq:sad}) are shown in (a) as guides to the eye. 
The fitting parameters are $\beta = 2.14$ and $\gamma = 0.0245$ for $\kappa = 0$, 
and $\beta = 3.4$ and $\gamma = 0.0054$ for $\kappa = 1$.
Analogous data for Model A are shown in Fig.~\ref{fig:a-sads} (Appendix \ref{sec:a_stability}).
\label{fig:b-fluctuations}}
\end{figure}

To estimate the relevance of the noise effects for Model B, 
the species abundance distributions (SADs) were investigated.
The SAD is the distribution of the populations of each species 
with binning on $\log_2$ scale, which is widely used in ecology. 
Figure \ref{fig:b-sad} shows the SAD for Model B with $\kappa = 0$ and $1$. 
Since the system without demographic stochasticity does not reach a stationary state, 
we divided the time series for $\kappa = 0$ into three regions; 
$r_1$, $r_2$, and $r_3$ are the regions 
where $t < 16$, $16 < t < 64$, and $t > 64$ million generations, respectively, 
in order to see how the distribution changes during the evolutionary process.
For $\kappa = 1$, the data are calculated for $t > 16$ million generations, 
where a statistically stationary state is realized. 
During the simulation, there are many species with small populations, which correspond to unsuccessful mutants. 
We filtered out unsuccessful mutants and obtained the ``core fixed-point communities'' 
by updating the population dynamics without the noise and mutations 
until all the fixed-point populations analytically calculated from the interaction matrix (Eq.~(\ref{eq:exact_sol}) ) 
became larger than the extinction threshold.
The SADs were calculated for these fixed-point communities.

The SADs for these communities are quite similar, while the peak position shows dependence on $\kappa$. 
The SAD for $\kappa = 1$ has a peak at higher population size than for $\kappa = 0$. 
The number of species with small populations are suppressed by the noise, 
and a small number of large-population species survive, 
which is a natural consequence of the law of large numbers. 
The difference in height corresponds to the difference in diversity between $\kappa = 0$ and $1$. 
The profile is well fitted by a function suggested by Pigollotti et al. \cite{PhysRevE.70.011916} 
\begin{equation}\label{eq:sad}
p(n) \propto \frac{e^{-\gamma n}}{ n^{1-\beta} },
\end{equation}
where $\gamma$ and $\beta$ are fitting parameters. 
This function interpolates between the well-known Fisher's log-series and the log-normal distributions.
The agreement of the fitting function for $\kappa = 0$ is reasonable although there is some difference at $n \sim 1$. 
For $\kappa = 1$, the fitting is reasonable only around the peak. 
The data have fatter tails than the function of Eq.~(\ref{eq:sad}). 

Since the noise term for the fixed-point community is of order $\kappa \sqrt{n} / 2$, 
the noise term is of the same order as $n^{\ast}$ if $n^{\ast} \lesssim 1$. 
Simply comparing with the SADs, which peak at $\gtrsim 10^2$; 
the noises are less than the corresponding population sizes for most of the species. 
However, if the system has weak linear stability, the noise may have relevant effects and cause extinctions of species. 
The eigenvalues of the linear stability matrix, $\lambda_{\mathbf{S}}$, are therefore estimated, 
and their probability density functions are shown in Fig.~\ref{fig:b-s-eigens}. 
The distributions for Model B show peaks slightly below $1$, 
which means the system is asymptotically stable, but the linear stability is weak. 
The distributions show dependence on $\kappa$, 
and the systems with $\kappa = 0$ are less stable than those with $\kappa = 1$. 
Thus, more stable communities are selected under the stochastic noise. 

We estimated how large the population fluctuations would be if the noise corresponding to $\kappa = 1$ 
were applied to the fixed-point communities. 
Following the discussion in \cite{zia-jpa}, 
we assume the probability that the system is found with a specific number of individuals $|n\rangle$ at a stationary state, 
${\cal P^{\ast}}(|n\rangle)$, takes the Gaussian form: 
\begin{equation}
	{\cal P}^{\ast}(|n\rangle) = (2\pi)^{-{\cal N}/2} (\det \mathbf{G})^{-1/2} 
	\exp{ \left[ - \frac{1}{2}  \langle \Delta n| \mathbf{G}^{-1}  |\Delta n\rangle \right] },
\end{equation}
where ${\cal N}$ is the number of resident species, $\mathbf{G}$ is the covariance matrix to be estimated, 
and $| \Delta n \rangle = |n\rangle - |n^{\ast} \rangle$ is the difference from the fixed point.
This approximation is valid only when the population fluctuations are small enough to neglect the nonlinearity of the population dynamics. 
Although this assumption is not satisfied for Model B, 
this discussion tells us that 
the population fluctuations could be of the same order as the population sizes and might cause extinctions. 
How the covariance matrix $\mathbf{G}$ is calculated is shown in Appendix \ref{sec:g-calculation}. 
Here we only show the distribution of the square root of the eigenvalues of $\mathbf{G}$, 
which correspond to the size of the population fluctuations.
The distribution is shown in Fig.~\ref{fig:b-g-eigens} in the same format as the SADs (binning in $\log_2$ scale). 
Comparing this figure with the SADs, 
the population fluctuations for $\kappa = 0$ would be of the same order as the population sizes.
Thus, the noise drives the species with little stability to extinction, 
and, as a result, cause the large decline in diversity.
The same linear stability analysis was also done for Model A (shown in Appendix \ref{sec:a_stability}).
Model A shows stronger linear stability and smaller eigenvalues of the covariance matrix $G$. 
Hence Model A is robust against the noise and does not show notable dependence on the stochastic noise. 

\begin{figure}[ht!]
\begin{center}
\subfigure{
\includegraphics[width=8.0cm]{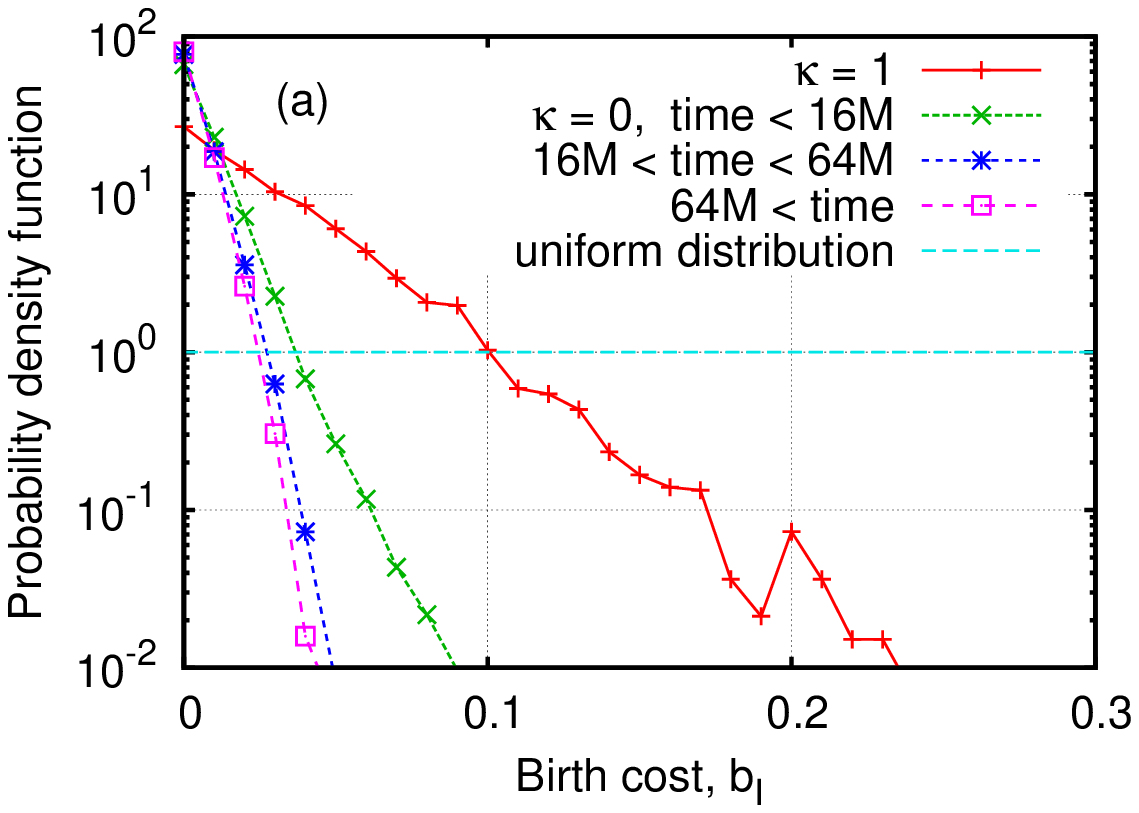}
\label{fig:b-b-histo}
}
\subfigure{
\includegraphics[width=8.0cm]{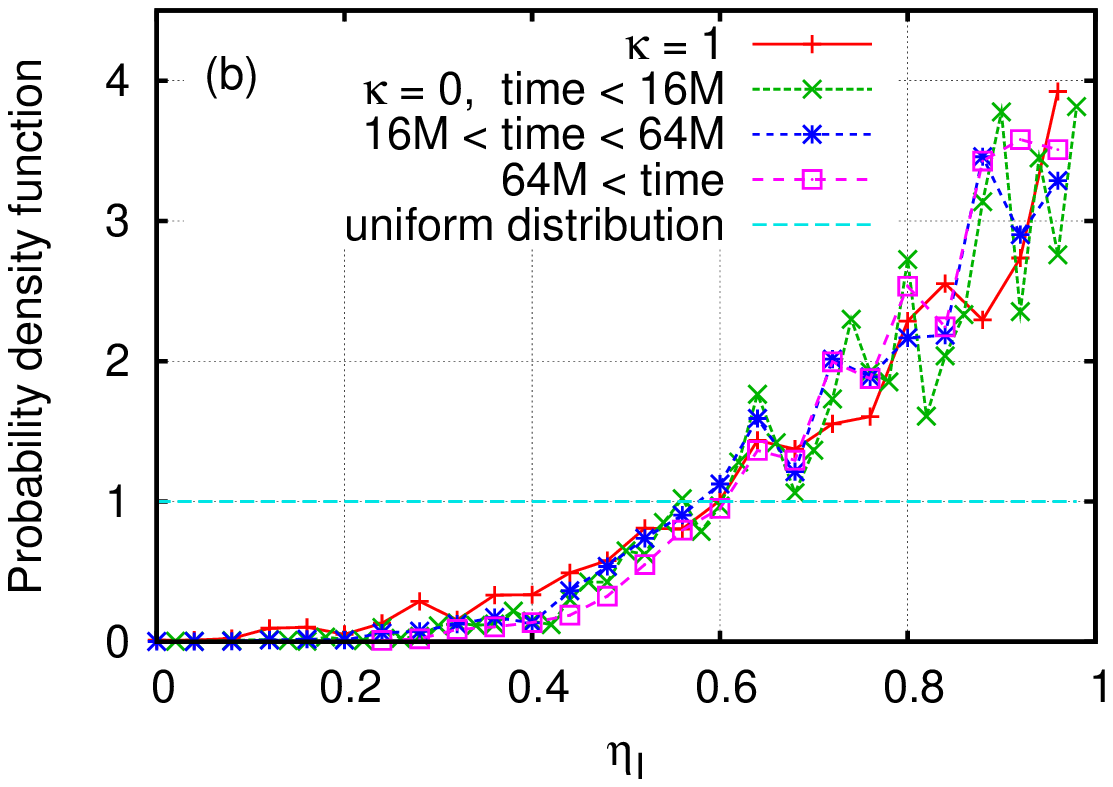}
\label{fig:b-eta-histo}
}
\subfigure{
\includegraphics[width=8.0cm]{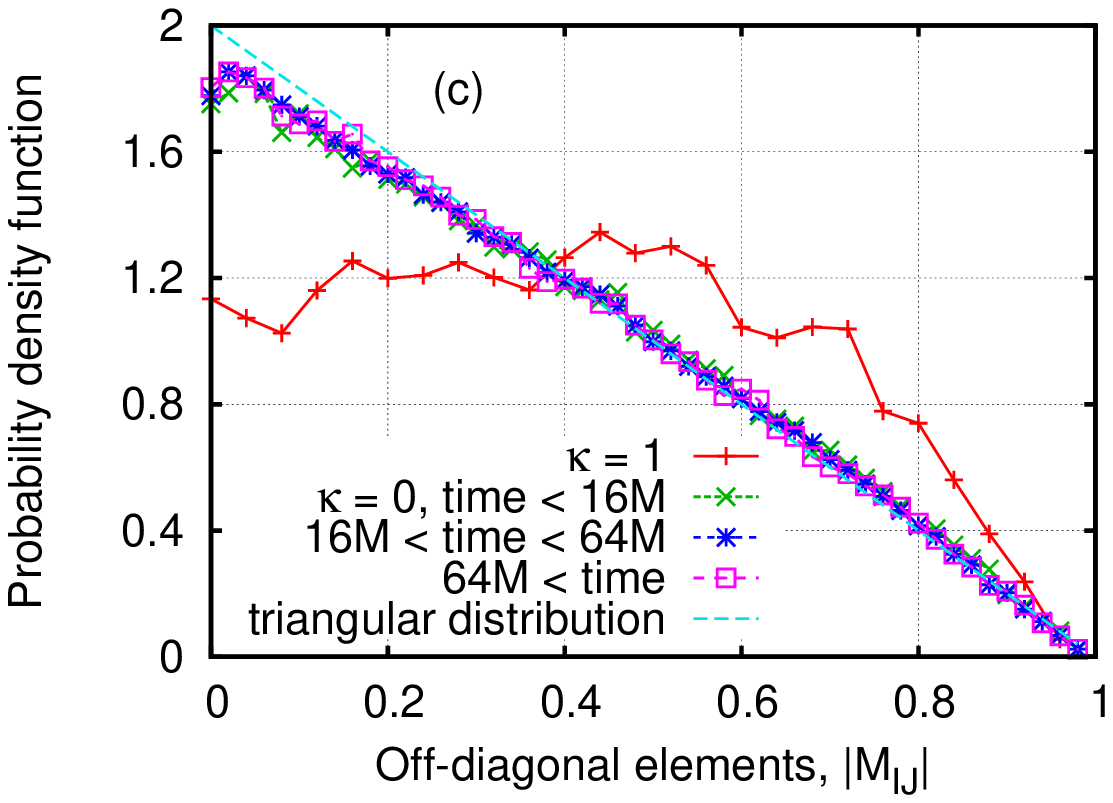}
\label{fig:b-mij-histo}
}
\end{center}
\caption{
(Color online) Probability density function of (a) birth cost, $b_I$, (b) coupling constant to resource, $\eta_I$, 
and (c) interaction matrix elements, $|M_{IJ}|$, for fixed-point communities of Model B. 
The data are obtained for $\kappa=1$ and $\kappa = 0$.
\label{fig:b-elements-pdf}}
\end{figure}

The next question is how the communities obtained for Model B become sensitive to the noise. 
The distribution of the birth cost $b_{I}$, the resource-coupling  coefficient $\eta_{I}$, 
and the interspecies interaction coefficient $M_{IJ}$ are shown in Fig.~\ref{fig:b-elements-pdf}.  
The distributions of $b_{I}$ and $M_{IJ}$ show dependence on $\kappa$, 
while those of $\eta_I$ do not show notable $\kappa$-dependence. 
In the communities which have evolved via population dynamics without demographic stochasticity, 
species with quite low $b_{I}$ are selected, while the selection on $M_{IJ}$ is weak.
The ratio of low $b_{I}$ species increases as the evolution proceeds. 
On the other hand, under the noise, 
the distribution of $b_{I}$ is not as extreme as for $\kappa = 0$, 
but the ratio of species with large $M_{IJ}$ becomes larger. 
Hence, 
the selection pressure is applied on the birth cost for $\kappa = 0$ 
while it is applied on the interspecies couplings for $\kappa = 1$. 

We also modified Model B so that all the species have the same birth cost, $b_{I}$ ($=0.1$). 
The results are shown in Appendix \ref{sec:b-b-const}. 
For this modified model, qualitatively similar results as for the original Model B are obtained: 
large decline in diversity, more strongly coupled communities, and stronger linear stability around the fixed points 
are observed under the noise.

\subsection{Long-term fluctuations}\label{subsec:long-term}
Not only the mean value of the diversity, but also its fluctuations during the evolution 
are affected by the demographic population fluctuations. 
The power spectral densities (PSDs) of the time series of diversity and total population size,
as well as probability densities of species lifetimes and QSS durations,
were calculated in order to evaluate the intermittency quantitatively. 
These results are of particular interest in connection with the dynamics of mass extinctions on geological time scales\cite{newman2003me}.

Power laws are estimated by fitting to $\log_2$-binned densities.
However, exponents obtained using the estimators from \cite{clauset:661} are not qualitatively different. 

\subsubsection{Model A}\label{subsubsec:long-term-modelA}
\begin{figure}[ht!]
\begin{center}
\subfigure{
\includegraphics[width=8.0cm]{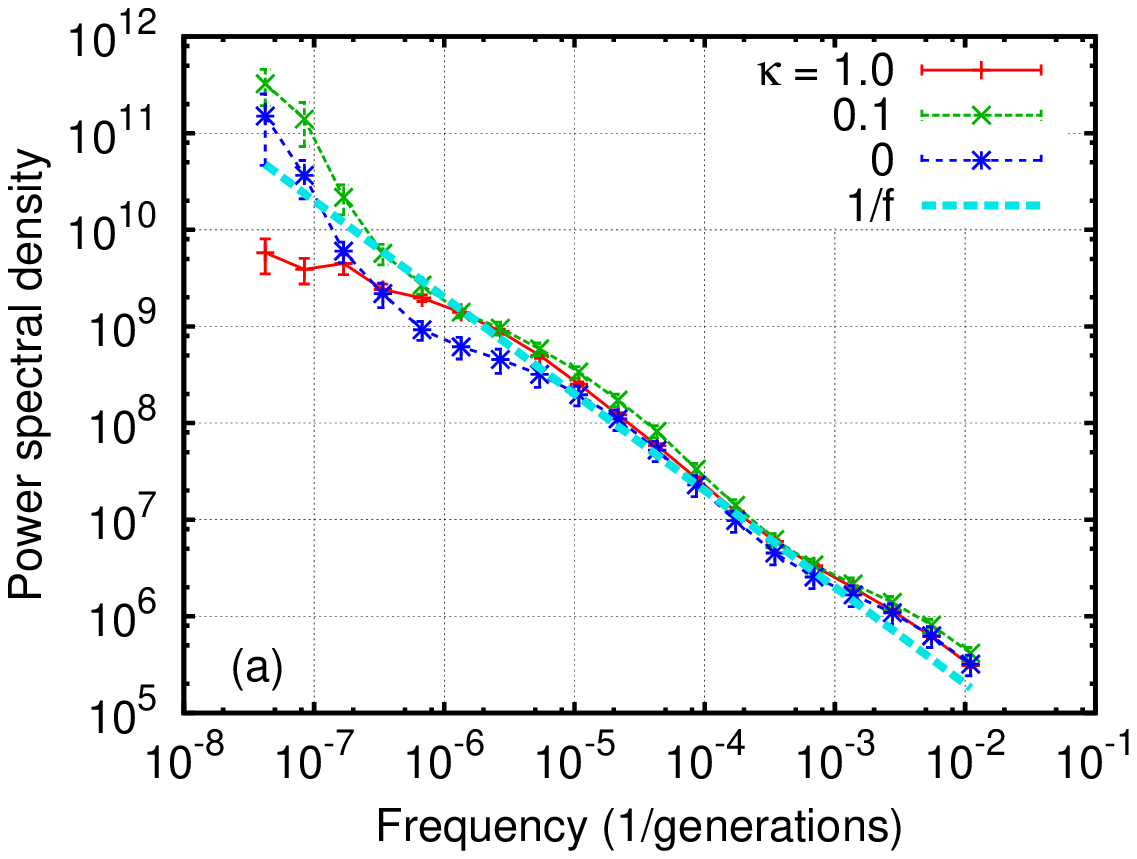}
\label{fig:tna-mut-psddiv}
}
\subfigure{
\includegraphics[width=8.0cm]{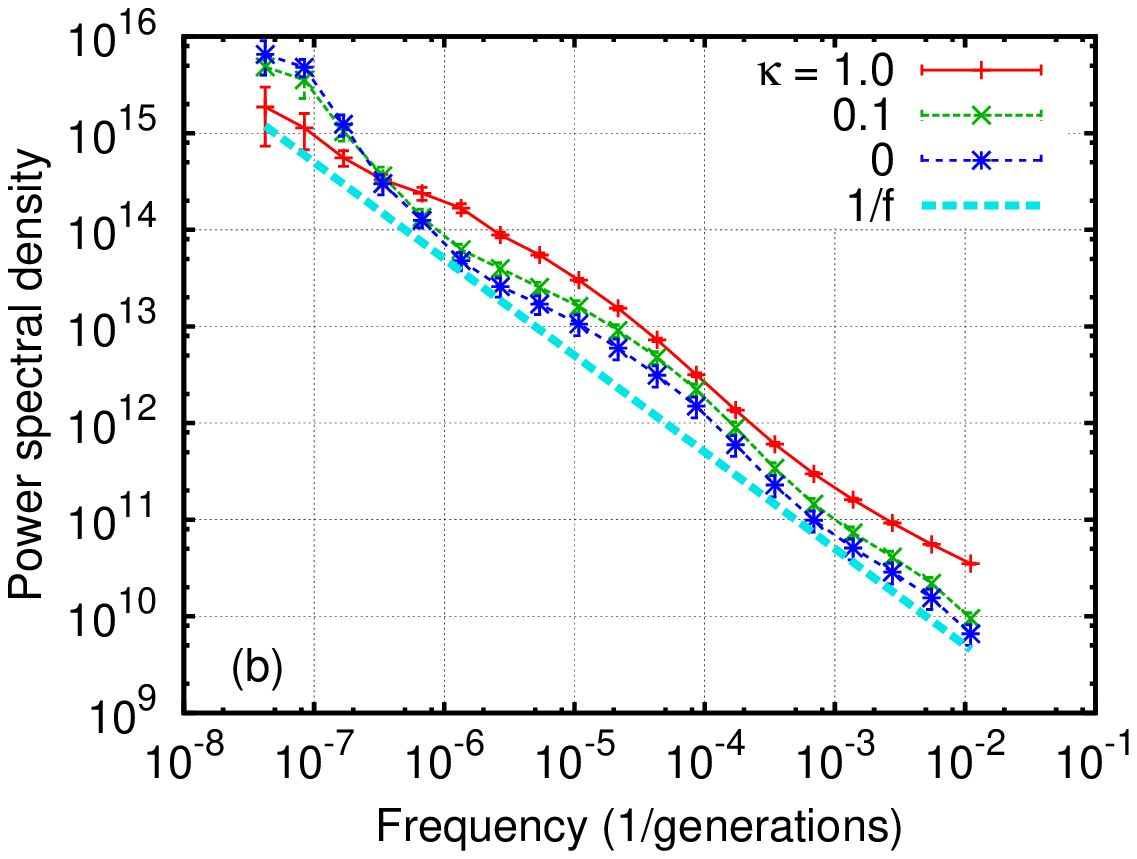}
\label{fig:tna-mut-psdbiomass}
}
\end{center}
\caption{
(Color online) PSDs of (a) exponential Shannon-Wiener diversities and (b) total population sizes 
for Model A with $\mu = 0.001$ at several noise levels.
In both figures, lines corresponding to $1/f$ are shown as guides to the eye.
Data are averaged over six independent runs, and their statistical errors are also shown.
\label{fig:tna-mut-psds}}
\end{figure}

We performed simulations of $2^{25} = 33554432$ generations with 
$2^{22} = 4194304$ generations as a ``warm-up'' period. 
This warm-up period is long enough to realize statistically stationary states. 
For each model and parameter, six independent runs were performed.
Figure \ref{fig:tna-mut-psds} shows the PSDs for several noise levels $\kappa$. 

\begin{figure}[ht!]
\begin{center}
\subfigure{
\includegraphics[width=8.0cm]{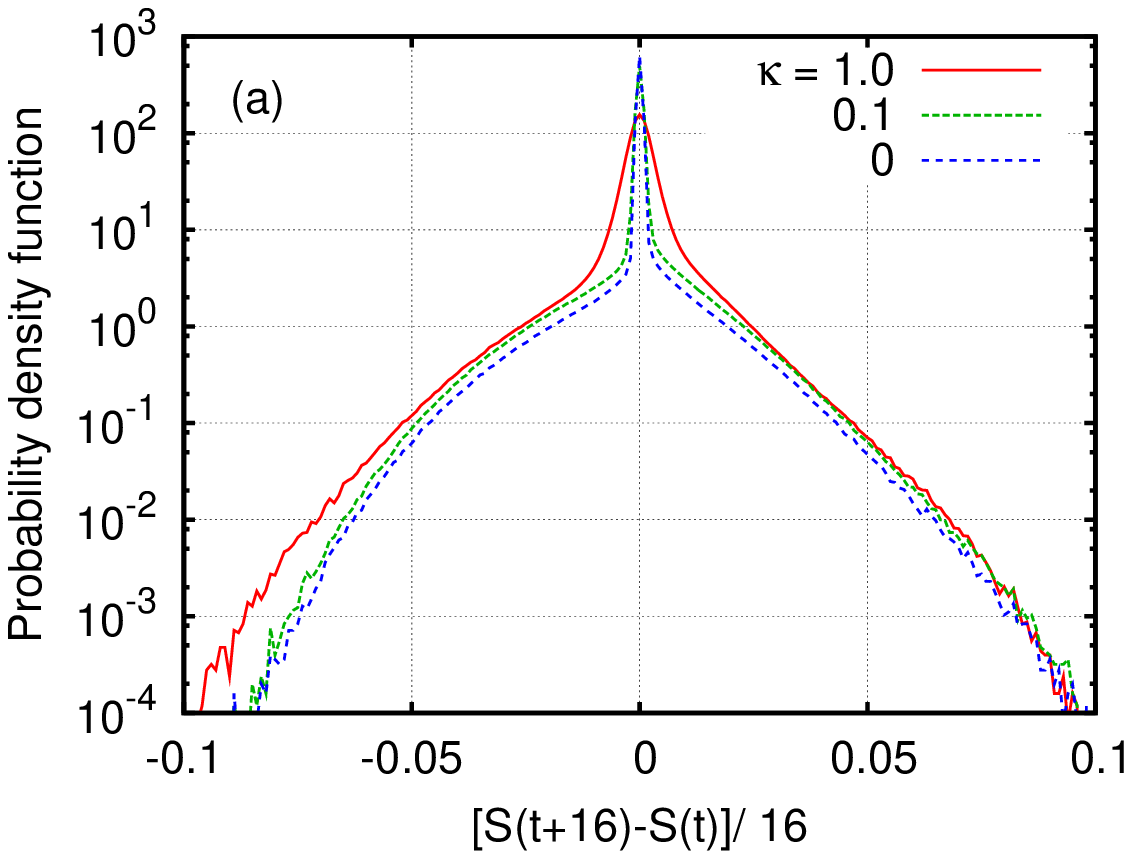}
\label{fig:a-mut-dSdt}
}
\subfigure{
\includegraphics[width=8.0cm]{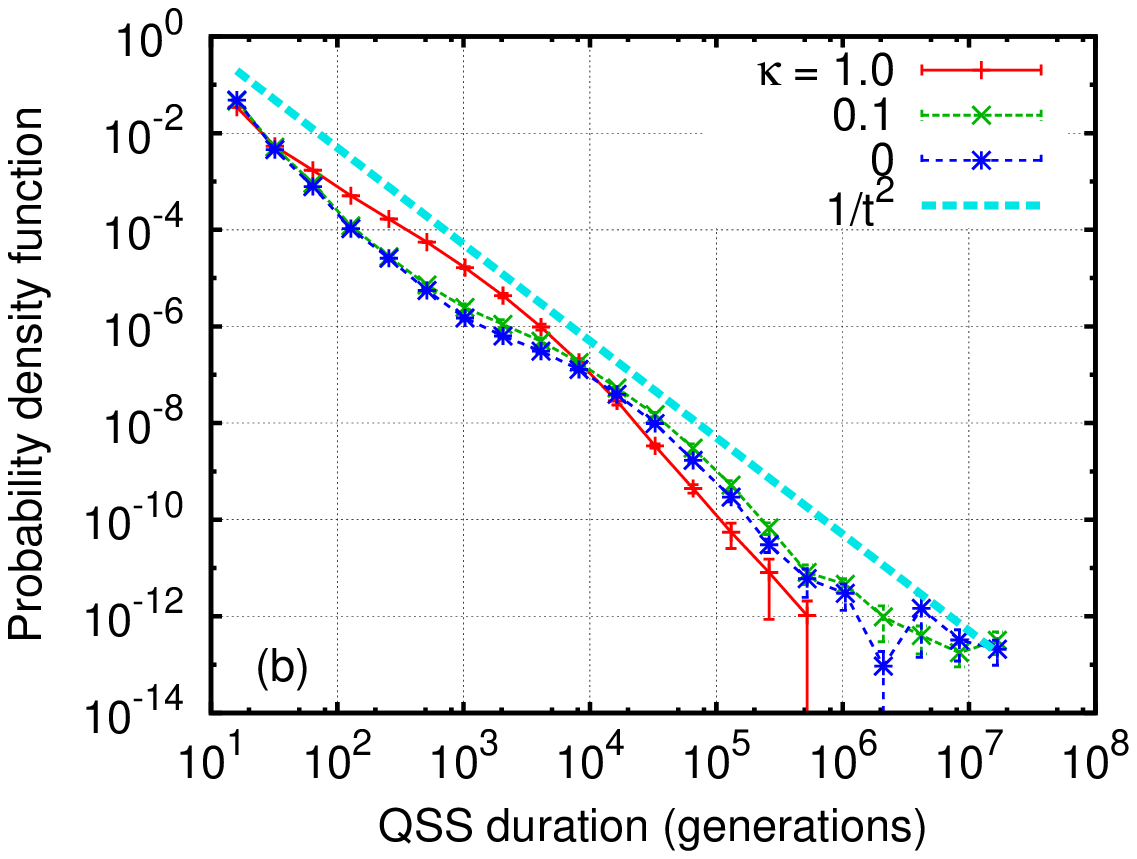}
\label{fig:a-mut-duration}
}
\end{center}
\caption{
(Color online) 
(a) The probability density functions of the logarithmic derivative of the diversity, $dS(t)/dt$, 
for Model A with $\mu = 0.001$ at several noise levels.
The data were averaged over 16 generations in each run, and then averaged over 6 independent runs.
(b) Log-log plot of the probability density functions of the duration of QSSs.
The QSSs are estimated as the periods between times 
when $|dS(t)/dt|$ exceeds a cutoff (here, $0.02$). 
The logarithmic derivative, $dS(t)/dt$, was averaged over 16 generations as in (a).
The line corresponding to a $t^{-2}$ power law is shown as a guide to the eye.
}
\label{fig:a-mut-dSdt-duration}
\end{figure}

\begin{figure}[ht!]
\begin{center}
\includegraphics[width=8.0cm]{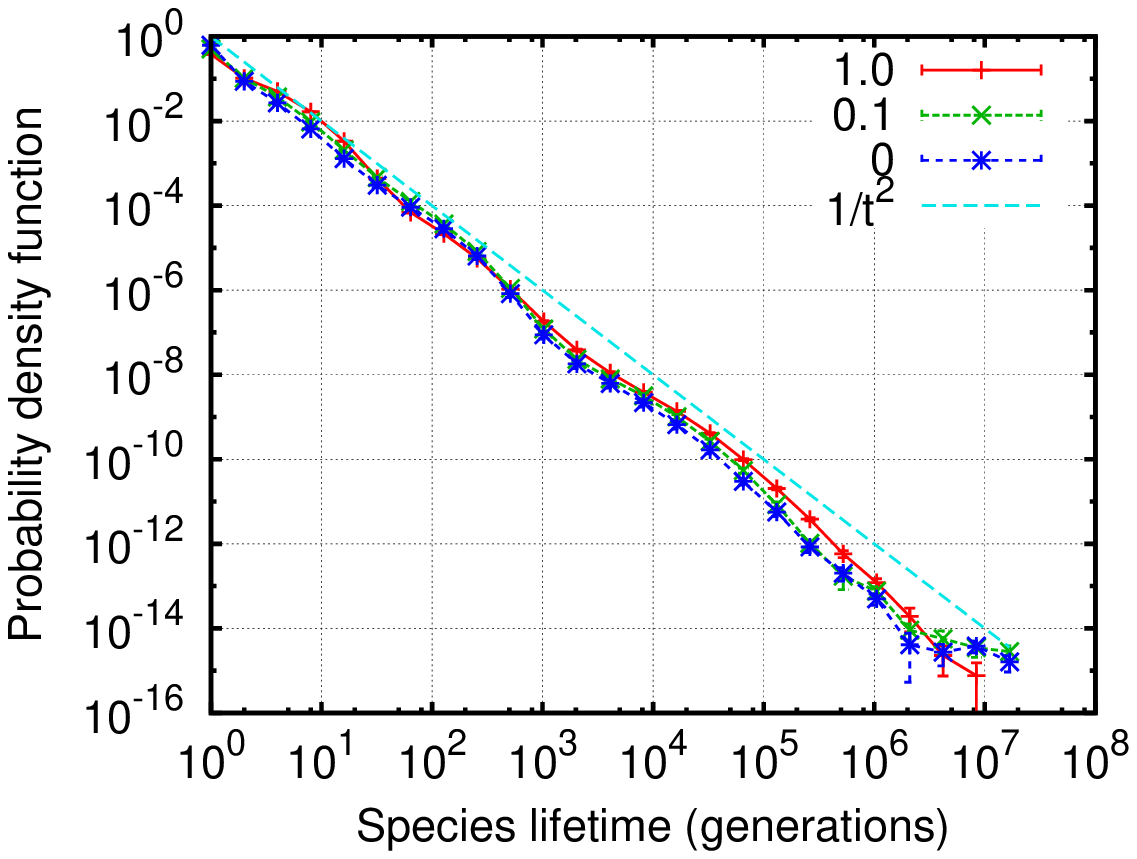}
\end{center}
\caption{
(Color online) Species-lifetime distributions plotted on log-log scale 
for Model A with $\mu = 0.001$ and $L=13$ at several noise levels.
The line corresponding to $t^{-2}$ is shown as a guide to the eye.
}
\label{fig:a-mut-lspan}
\end{figure}

For Model A, both diversity and total population size generally show approximate $1/f$ fluctuations 
for all values of the noise level $\kappa$. 
The PSDs for weak population fluctuations show approximate $1/f$ power-law behavior over more than five decades.
Thus the $1/f$ fluctuations found in the individual-based Model A \cite{rikvold2003pea} 
are robustly reproduced, even with deterministic population updates. 
Under very strong population fluctuations, 
the possibility of extinctions caused by the population fluctuations is not negligible, 
and few communities are able to persist over very many generations. 
As a result, the PSDs for high $\kappa$ are not $1/f$ like at very low frequencies. 
This effect of population fluctuations is also observed in the QSS duration distributions and the species lifetime distributions 
as seen in Figs.~\ref{fig:a-mut-duration} and \ref{fig:a-mut-lspan}. 

Figure \ref{fig:a-mut-dSdt} shows the logarithmic derivative of the time series of the diversity (i.e., $dS/dt$), 
which is averaged over 16 generations.
Each curve has a sharp peak around the center and relatively wide wings in both tails. 
The sharp peak around zero represents that the community is in a quiet period. 
The small diversity fluctuations arise from the population fluctuations of coexisting species 
and the repetitive emergence and extinction of unsuccessful mutants. 
On the other hand, the large wings represent large rearrangements of the species composition. 
By measuring $dS/dt$, we can judge whether the system is in a quiet period 
where the species composition remains approximately constant, 
or in an active period where the dominant species are replaced rapidly. 
This type of profile is similar to that of the corresponding individual-based Model A \cite{rikvold2003pea,Rikvold:2007lr}. 

The duration distributions for QSSs are shown in Fig.~\ref{fig:a-mut-duration}. 
The QSSs are estimated as the periods between times when $|dS/dt|$ exceeds a cutoff (here, $0.02$).
The distributions show approximate $1/t^2$ power laws, 
reproducing the result for the corresponding individual-based model. 

Figure \ref{fig:a-mut-lspan} shows 
the species-lifetime distributions at several noise levels.
The distribution is fitted by a power law over more than six decades, 
and the exponent of the observed distribution is about $2.2$, 
which is in reasonable agreement with the individual-based Model A \cite{rikvold2003pea}. 
The distribution does not show important dependence on the population fluctuations.

\subsubsection{Model B}\label{subsubsec:long-term-modelB}

Next we show the results for Model B. 
We performed six independent runs of $2^{26} = 67108864$ generations 
with $2^{24} = 16777216$ generations as a warm-up period. 
The mutation rate $\mu=0.0005$ and the genome length $L=18$ were used.

Although the species populations basically fluctuate around their fixed-point values, 
the probability that a species population touches the extinction threshold increases 
under strong stochastic population fluctuations. 
Therefore, both diversity and total population size tend to decrease as $\kappa$ increases. 
We also note that the diversity for $\kappa=1$ is approximately the same as that of the individual-based Model B \cite{rikvold2007ibp}, 
indicating that the current model with $\kappa=1$ is a good approximation to the individual-based one. 
The fluctuations of the diversity for smaller $\kappa$ are also larger than for larger $\kappa$.
These fluctuations for small $\kappa$ come from introductions of mutants and extinctions of species.
When $\kappa$ is small (e.g. $\kappa=0.1$ or $0$), the system shows high diversity and large population size, 
which are still growing even after $80$ million generations. 
In addition, the fluctuations are less intermittent than for $\kappa=1$.
This intermittency will be quantitatively estimated below.
		
\begin{figure}[ht!]
\begin{center}
\subfigure{
\includegraphics[width=8.0cm]{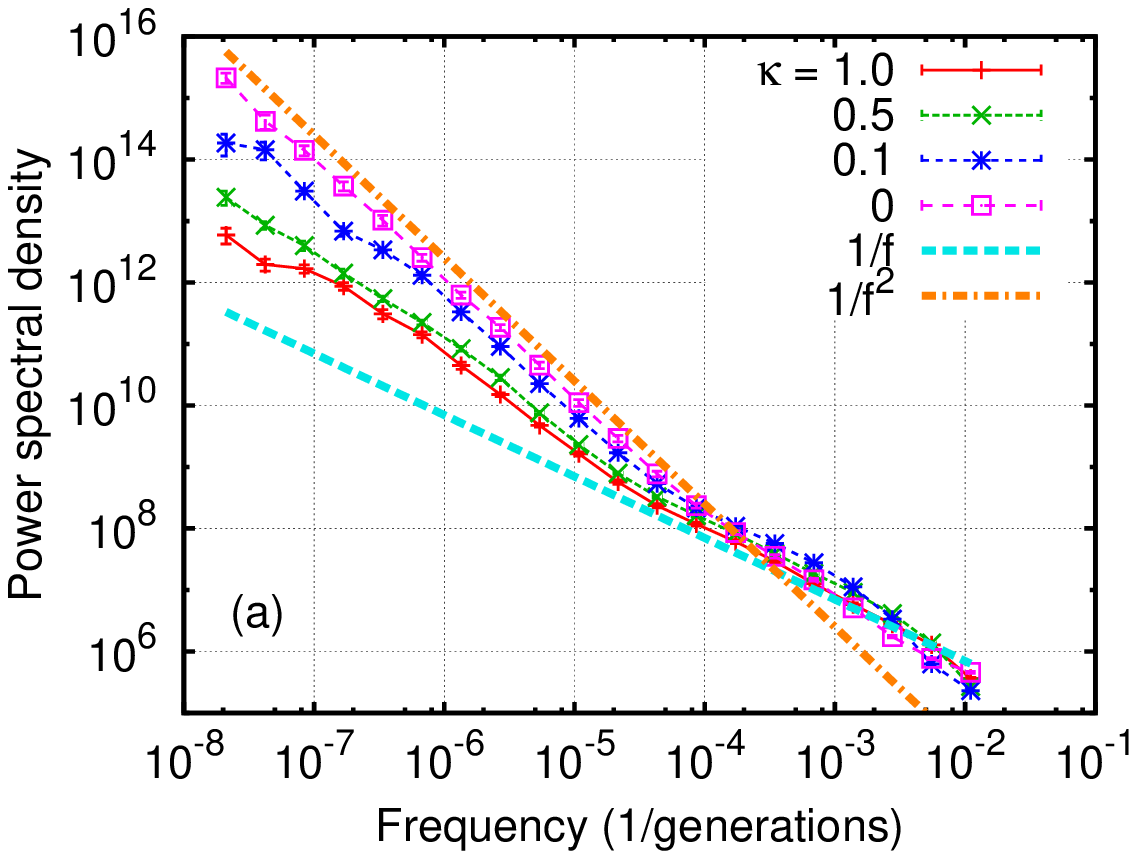}
\label{fig:b-mut-psd-div}
}
\subfigure{
\includegraphics[width=8.0cm]{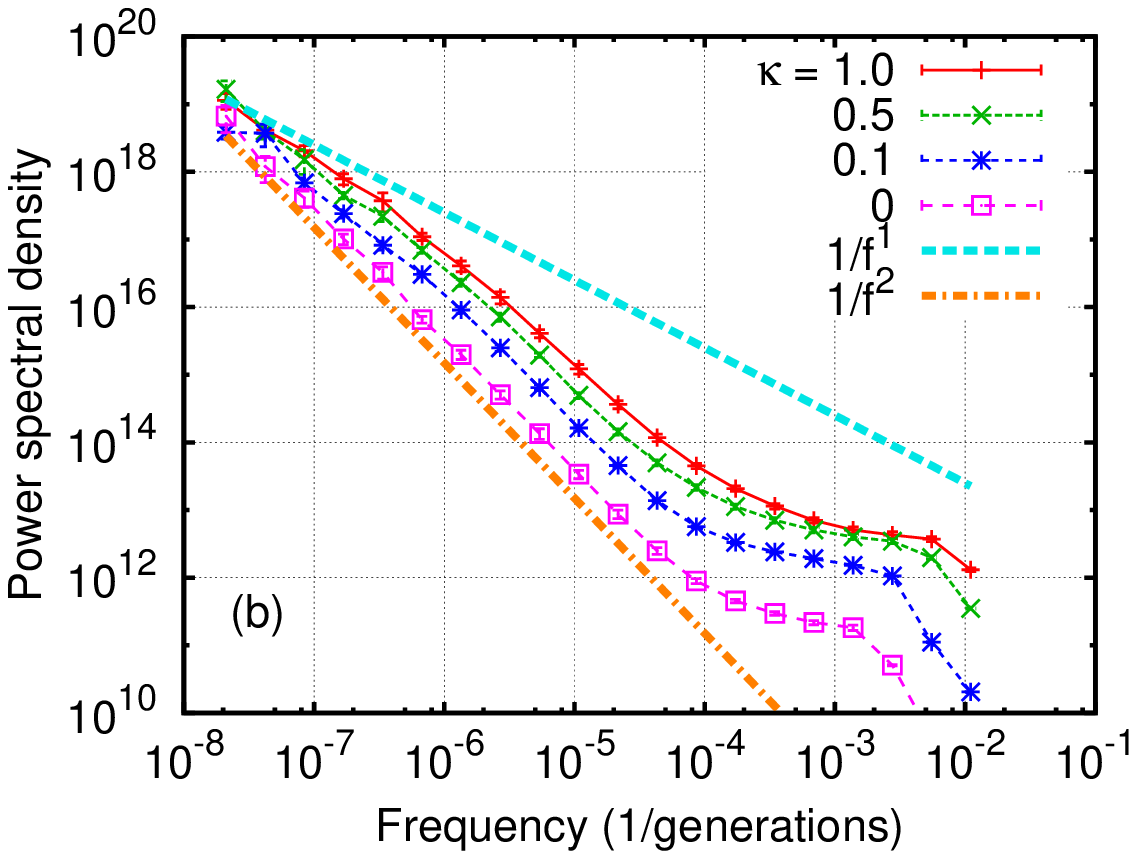}
\label{fig:b-mut-psd-biomass}
}
\end{center}
\caption{
(Color online) PSDs of (a) exponential Shannon-Wiener diversities and (b) total population sizes 
for Model B with $\mu = 0.0005$ at several noise levels. 
The data are averaged over six independent runs and their (small) statistical errors are also shown.
The straight lines in each figure represent $1/f^{\alpha}$ power laws with exponents 
$\alpha = 1$ and $2$ as guides to the eye. 
The shoulder in the population-size PSD at high frequencies is due to self-excited population oscillations. 
\label{fig:b-mut-psds}}
\end{figure}

Figure \ref{fig:b-mut-psds} shows PSDs of the diversities and the total population sizes 
at several noise levels. 
The PSDs of both diversity and total population size show power laws. 
The exponent depends on $\kappa$. 
When $\kappa$ is large, the PSDs show approximate $1/f^{\alpha}$ behavior with $\alpha=1.3 \sim 1.4$. 
This exponent $\alpha$ is in reasonable agreement with the individual-based Model B \cite{rikvold2007ibp} 
although it is slightly larger. 
As $\kappa$ decreases, the exponent gets closer to $2$; 
indicating that the diversity and the total population size both fluctuate like random walks.
Thus the population fluctuations change not only the average value of the diversity 
but also its fluctuations on evolutionary time scales.
In the individual-based model, 
the evolution proceeds intermittently, repeating quiet periods punctuated by brief active periods.
However, in the deterministic model, 
the evolution proceeds rather gradually, and the community composition changes continuously. 

Since the stationary state is not realized on the time scale we observed for $\kappa = 0$, 
we also calculated the PSDs of a corrected time series. 
First we calculated least-squares fits for the time series, 
and then calculated PSDs of the difference of the time series from the linear fit. 
The result (not shown) does not show notable differences from Fig.~\ref{fig:b-mut-psds}. 

\begin{figure}[ht!]
\begin{center}
\subfigure{
\includegraphics[width=8.0cm]{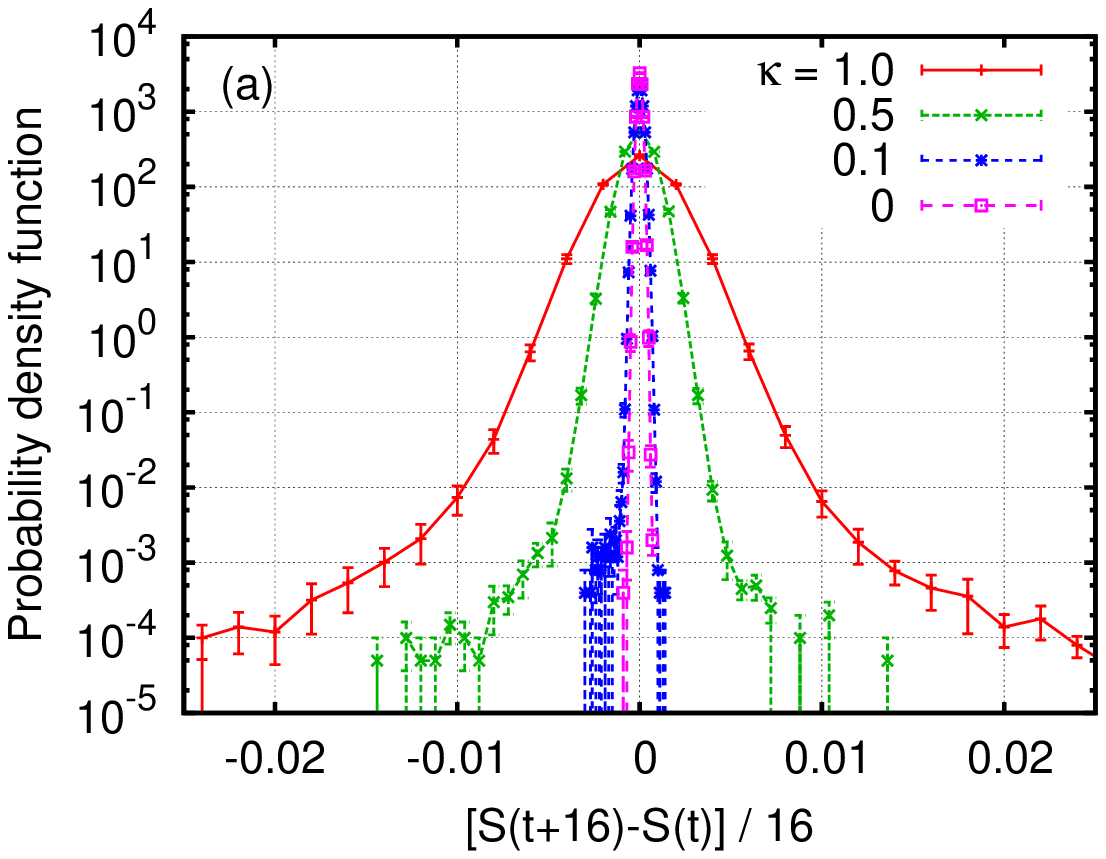}
\label{fig:b-mut-dsdt}
}
\subfigure{
\includegraphics[width=8.0cm]{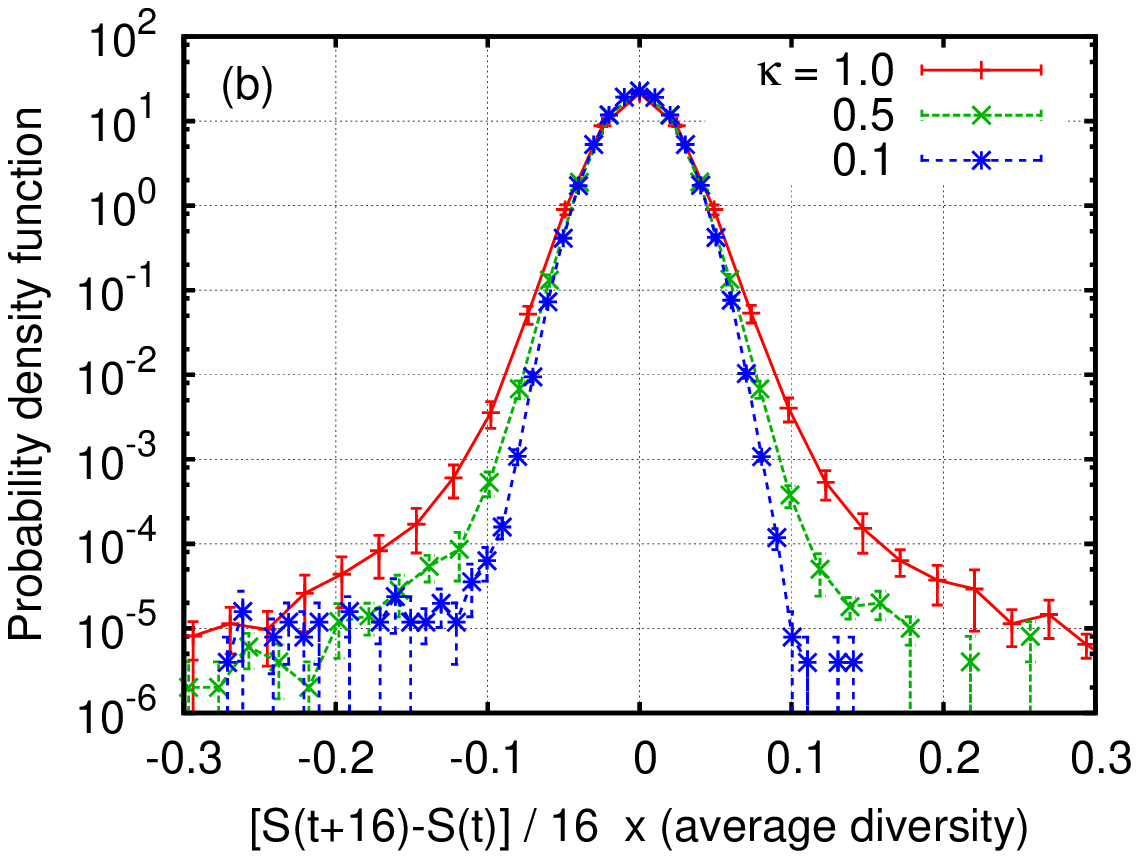}
\label{fig:b-mut-dsdt-normed}
}
\end{center}
\caption{
(Color online) (a) Probability density functions of the logarithmic derivative of the diversity $[S(t+16)-S(t)]/16$ 
for Model B with $\mu = 0.0005$ at several noise levels. 
(b) Probability density functions of $\{[S(t+16)-S(t)] / 16 \} \times \bar{D}$, where $\bar{D}$ is the average diversity.
The data are averaged over six independent runs and their statistical errors are also shown.
\label{fig:b-mut-dsdts}}
\end{figure}

Figure \ref{fig:b-mut-dsdt} shows the probability density function of 
the logarithmic diversity derivative, $dS/dt$. 
Since the averaged diversity for smaller $\kappa$ is much higher than for larger $\kappa$,
the distribution for smaller $\kappa$ is quite sharp. 
This is due to the high diversity realized for weak noises. 
To eliminate this effect, the derivatives normalized by the average diversities are shown in Fig.~\ref{fig:b-mut-dsdt-normed}. 
We calculated $[S(t+16)-S(t)] / 16 \times \overline{D}$ for every $16$ generations, 
where $\overline{D}$ is the average diversity. 
Therefore the $x$-axis of Fig.~\ref{fig:b-mut-dsdt-normed} has the dimension of $({\rm diversity})\cdot({\rm time})^{-1}$.
The sharpness of this normalized data are almost similar, 
therefore the absolute diversity fluctuations are only weakly $\kappa$-dependent. 

The distribution for $\kappa=1$ has a Gaussian center and large wings 
and looks similar to the individual-based Model B \cite{rikvold2007ibp}.
However, the distributions for smaller $\kappa$ look different.
When $\kappa=0.1$, 
the distribution is quite well fitted by a Gaussian distribution without wings \cite{zia-jpa}. 

\begin{figure}[ht!]
\begin{center}
\includegraphics[width=8.0cm]{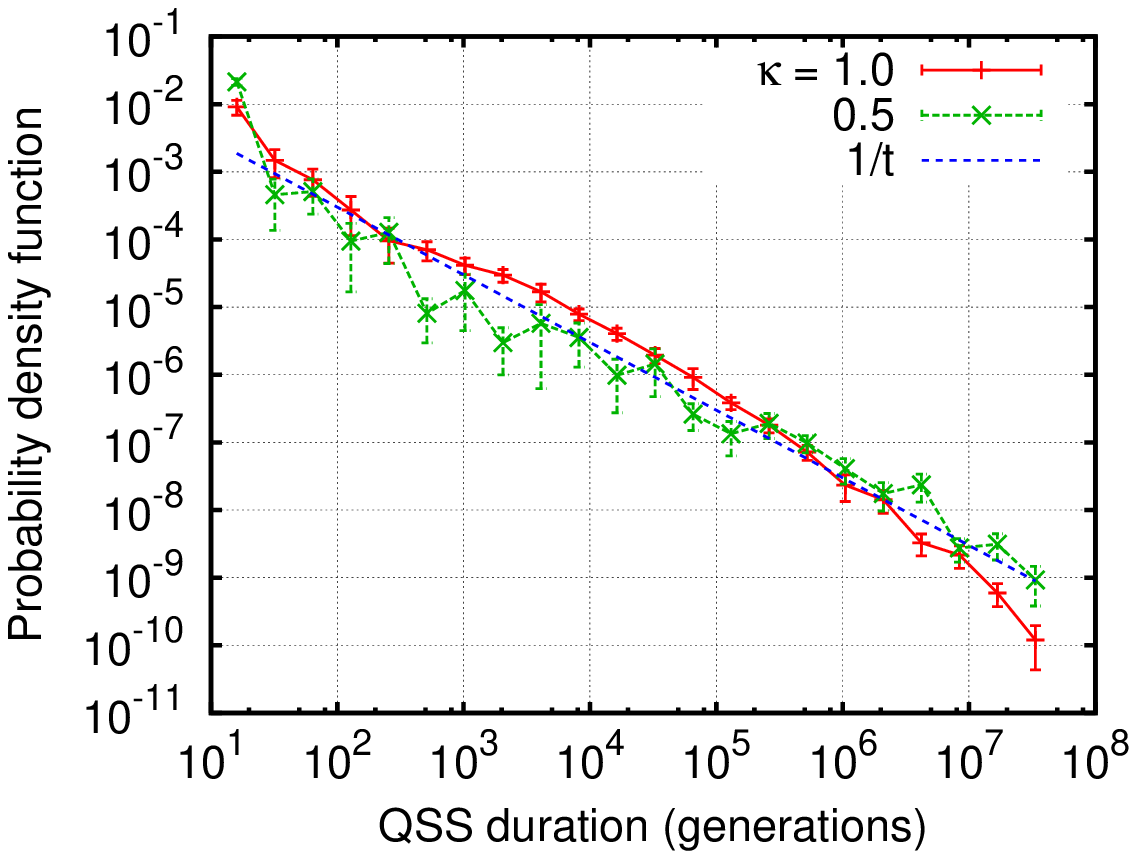}
\end{center}
\caption{
(Color online) Probability density functions of QSS duration for Model B with several values of $\kappa$.
The QSSs are estimated as the periods between times when $|dS(t)/dt| \times \bar{D}$ exceeds a cutoff (here, 0.1).
The line corresponding to $1/t$ is shown as a guide to the eye.
}
\label{fig:b-mut-duration}
\end{figure}

The distributions of QSS durations are calculated and shown in Fig.~\ref{fig:b-mut-duration} 
for $\kappa=1$ and $0.5$.
The QSSs are estimated in the same way as the previous model, 
but different thresholds are used for each $\kappa$ because the profiles show large dependence on $\kappa$.
The threshold for estimating QSS is $0.1/\overline{D}$, where $\overline{D}$ is the average diversity. 
The distributions show approximate $1/t$ power laws regardless of the noise level.
This is consistent with the original individual-based Model B \cite{rikvold2007ibp}
in which a $t^{-1}$ power law is observed in the QSS duration distribution \cite{Rikvold:2007lr,rikvold2007ibp}. 
For $\kappa = 0.1$ and $0$, it is impossible to estimate QSS durations 
since the distribution of logarithmic derivatives of the diversity for these parameters does not have large wings. 
If we estimate the QSS with a threshold which corresponds to the Gaussian region, 
clear exponential decay is observed. 
Hence the small fluctuations occur randomly and do not have remarkable long-time correlations.

\begin{figure}[ht!]
\begin{center}
\includegraphics[width=8.0cm]{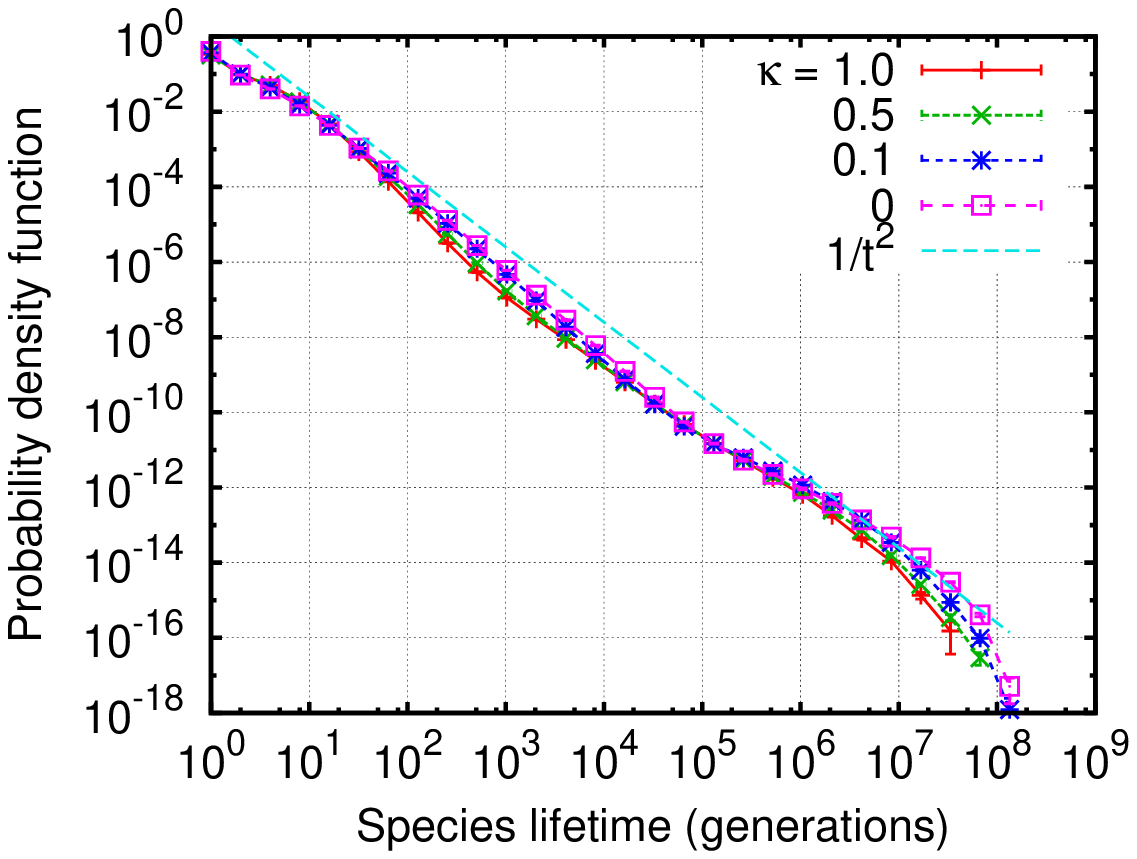}
\end{center}
\caption{
(Color online) Species-lifetime distributions plotted on log-log scale 
for Model B with $\mu = 0.0005$ at several noise levels.
The line corresponding to $t^{-2}$ is shown as a guide to the eye.
}
\label{fig:b-mut-lspan}
\end{figure}

Species-lifetime distributions for several $\kappa$ are shown in Fig.~\ref{fig:b-mut-lspan}. 
The distributions show a reasonable fit to a $t^{-2}$ power law for every $\kappa$. 
The average species lifetime for larger $\kappa$ is slightly less than for smaller $\kappa$, 
but the dependence on the noise strength is slight.
Hence the lifetime distribution is not notably affected at the species level even for small $\kappa$, 
while it is affected at the community level for $\kappa \leq 0.1$.

\section{Summary and Discussion}\label{sec:summary}

The effects of demographic stochasticity are explored for two types of biological macro-evolution models. 
The demographic stochasticity is modeled by the noise term of the population dynamics. 
For the mutualistic communities obtained by Model A, 
the noise does not have an important effect, 
and the deterministic description does not alter the picture obtained for the corresponding individual-based model. 
On the other hand, 
the predator-prey model (Model B) shows a remarkable decline in diversity at higher noise levels.
This is because the deterministic population dynamics allow species to coexist with low linear stability, 
which are pushed into extinction in the stochastic population dynamics. 
Without the noise, the distribution of the birth cost $b_{I}$ has a sharp peak close to zero, 
while the distribution of the coupling constants $M_{IJ}$ is almost the same as the original distribution. 
With moderate noise, the selection pressure for small $b_{I}$ becomes less extreme, 
and species with larger $M_{IJ}$ are selected. 
Hence, strongly coupled communities are selected under the noise, 
while species with low birth costs are selected at low noise levels.

For model B, species must have 
a strong coupling to the external resource 
or strongly predatory interaction coefficients compared to the birth cost to sustain their populations. (See Eq.~(\ref{eq:delta-modelB}).)
Our result that communities with larger $M_{IJ}$ are selected under the noise 
looks contradictory to the classical consensus on the relation between stability and complexity:
a community tends to be less stable when the interactions are dense and strong 
\cite{GARDNER:1970fj,may72:_will_large_compl_system_be_stabl,robert74:_stabil_of_feasib_random_ecosy,TREGONNING:1979yq,roberts80:_robus_of_natur_system,mccann-2000}.
This apparent contradiction is due to the antisymmetric correlation of the linear stability matrix. 
The off-diagonal parts of the linear stability matrix in Model B is closer to an antisymmetric form. 
The ratio 
\begin{equation}
  \label{eq:ls-symmetry}
  \frac{\Lambda_{IJ}}{\Lambda_{JI}} = \frac{ n_{I}^{\ast} (M_{IJ}-b_{I})}{ n_{J}^{\ast} (M_{JI}-b_{J})} 
\end{equation}
often becomes negative since the interaction matrix $M$ is antisymmetric. 
Eigenvalues originating from antisymmetric off-diagonal matrix elements are all pure imaginary, 
therefore the large $M_{IJ}$ do not destabilize the system significantly. 
Actually, the distribution of the diagonal parts $\Lambda_{II}$ is similar to the distribution of eigenvalues,
which implies that contributions of the off-diagonal parts are not critical.
Moreover, if the average predation rate is large compared to the birth cost, 
species tend to have larger equilibrium populations.
That makes the species have higher resistance against the demographic noise. 
Thus, large $M_{IJ}$ often leads to larger equilibrium populations without sacrificing the linear stability. 
We speculate that such selection of stronger coupling interactions 
occurs in a wide class of predator-prey population dynamics models under demographic noise. 

The dynamics on evolutionary time scales for Model B is also altered by the noise. 
When an appropriate amount of noise is applied, the system shows approximate $1/f$ fluctuations in the evolutionary dynamics. 
The time series consist of long quiet periods, during which the species compositions are steady, 
and short active periods, in which rearrangements of species compositions occur with relatively large-scale extinctions. 
The duration distribution for the quiet periods is an approximate power law, and $1/f$ fluctuations are found for the diversity index 
and the total population size. 
However, in the limit of no demographic stochasticity, 
this intermittent dynamics is replaced by a more gradual one, 
and the time series of the diversity index and the population size become Ornstein-Uhlenbeck processes. 
As the noise increases, the $1/f^2$ PSDs gradually change toward $1/f$ fluctuations.
We speculate that this is due to the smallness of the mutant's population and the weak linear stability.
Since mutants are quite prone to go extinct under the demographic noise,
QSS communities are more robust against the invasions of new species.
Similar effect is also reported in another model \cite{claessen2007delayed}.


The results shown in this paper indicate that 
models without noise may be remarkably different from models with noise. 
Without demographic noise, communities with weak stability that would be destroyed under the noise can emerge. 
Since the noise effect can be more important than suggested by a naive $1/\sqrt{N}$ prediction, 
the relevance of demographic stochasticity is not limited to small-scale communities, 
such as isolated islands, lakes, and experimental situations in microbiology, 
but can also exist for larger-scale ecosystems.

\section*{Acknowledgements}
This work was partly supported by 
21st Century COE Program ``Applied Physics on Strong Correlation''
from the Ministry of Education, Culture, Sports, Science, and Technology of Japan, the JSPS (Grant No. 19340110), 
and GRP of KAUST (Grant No. KUK-I1-005-04). 
Y.M. appreciates hospitality at Florida State University, 
where work was supported by U.S. NSF Grants No. DMR-0444051 and No. DMR-0802288.

\appendix
\section{Stochasticity in the number of offsprings}\label{sec:f-fluc}
It is straightforward to extend the model so that the number $F$ of offspring per individual becomes random.
Suppose the probability density function (pdf) of $F$, $q(F)$, is approximated 
by a Gaussian distribution with mean $\mu_{F}$ and variance $\sigma_{F}^2$.
The pdf of the number of individuals in the next generation, born to parents of species $I$, 
$p_I(n')$ for $n' > n_{\rm thr}$ is then given by
\begin{equation}
  p_I(n') = \int_{n_{\rm thr}}^{\infty} \mathcal{N}[n_{I}P_{I}, n_{I}P_{I}(1-P_{I})](x) \times \mathcal{N}[ x \mu_F, x \sigma_{F}^2]( n') dx,
\end{equation}
where $\mathcal{N}[\mu, \sigma^2](x)$ is a Gaussian distribution with mean $\mu$ and variance $\sigma^2$.
Thus, the fluctuations in population dynamics are more enhanced when $F$ fluctuates.
The limit $\sigma_{F} \rightarrow 0$ corresponds to the model considered in the body of this paper.

\section{Calculation of the fixed point and the linear stability}\label{sec:fp-calculation}
We briefly show this solution for the sake of completeness and readers' convenience, 
although it is shown in detail in \cite{rikvold2007ibp,Rikvold:2007lr}.
At the fixed point, the condition $| P(R,\{n^{\ast}\}) \rangle = 1/F$ is satisfied, 
where $|P\rangle$ is the column vector of the reproduction probabilities. 
Taking the logarithm of this equation gives rise to $\mathcal{N}$ linear equations, 
where $\mathcal{N}$ is the number of populated species. 
The solution for $|n^{\ast}\rangle$ is 
\begin{equation}
\label{eq:exact_sol}
|n^{\ast}\rangle = - \mathbf{ \hat{M}}^{-1} \left[ |\eta\rangle R - |\tilde{b}\rangle N_{\rm tot}^{\ast} 
- |1\rangle(N_{\rm tot}^{\ast})^2/N_0 \right],
\end{equation}
where $\mathbf{\hat{M}}^{-1}$, $|\eta\rangle$, $|\tilde{b}\rangle$, $|1\rangle$ are 
the inverse of the submatrix of $\mathbf{M}$ corresponding to the present species, 
and the column vectors of $\eta_I$, $b_I-\ln{(F-1)}$, and ones, respectively.
To find each $n_I^{\ast}$, we must first obtain $N_{\rm tot}^{\ast}$ ( $=\sum n_I^{\ast}$ ) as follows:
\begin{eqnarray}
N_{\rm tot}^{\ast} = \left\{
\begin{array}{ll}
  \frac{ \Theta N_0}{2} + \sqrt{ \left( \frac{\Theta N_0}{2} \right)^{2} + R \mathcal{E}N_0 } & (N_{0} \neq \infty) \\
  -R\mathcal{E}/\Theta & (N_0 = \infty)
\end{array} \right., 
\end{eqnarray}
where $\mathcal{E}$ and $\Theta$ are defined as 
\begin{equation}
\mathcal{E} = \frac{ \langle 1 | \mathbf{\hat{M}}^{-1} | \eta \rangle}{ \langle 1 | \mathbf{\hat{M}}^{-1} | 1 \rangle }
\end{equation}
and
\begin{equation}
\Theta = \frac{ 1- \langle1| \mathbf{\hat{M}}^{-1} | \tilde{b} \rangle }{ \langle1|\mathbf{\hat{M}}^{-1}|1\rangle },
\end{equation}
respectively.
The coefficients $\mathcal{E}$ and $\Theta$ can be considered as an effective coupling to the external resource 
and an effective interaction strength, respectively.
To find each $n_I^{\ast}$ separately, 
we now only need to insert this solution for $N_{\rm tot}^{\ast}$ in Eq.~(\ref{eq:exact_sol}).

Linear stability around fixed points can also be estimated analytically. 
The elements of the linear stability matrix $\mathbf{S}$ are
\begin{equation}
S_{IJ} =\frac{ \partial \left(n_I(t+1)\right) }{ \partial n_J(t)} 
=  \delta_{IJ} + \Lambda_{IJ},
\end{equation}
where $\delta_{IJ}$ is the Kronecker delta, and 
\begin{equation}
 \Lambda_{IJ} = 
(1-\frac{1}{F}) \frac{n_{I}^{\ast}}{N_{\rm tot}^{\ast}} 
\left[ M_{IJ} - \frac{ R\eta_{I} + \sum_{K}{M_{IK}n_K^{\ast}}}{N_{\rm tot}^{\ast}} - \frac{N_{\rm tot}^{\ast}}{N_0} \right]
\end{equation}
is the community matrix.
The system is stable against perturbations when all the eigenvalues of $\mathbf{S}$ are less than unity in magnitude.

\section{Calculation of the covariance matrix $G$}\label{sec:g-calculation}

For the calculation of  the covariance matrix $\mathbf{G}$, 
we need not only the stability matrix $\mathbf{S}$ but also the noise matrix $\mathbf{H}$, 
which is defined as the covariance matrix of the noise term of the population dynamics. 
Since the noise for each species is independent, $\mathbf{H}$ is a diagonal matrix and is written as 
\begin{eqnarray}
	H_{IJ} &=& \delta_{IJ} F^2 \kappa^2 n_I^{\ast} P_I(|n_I^{\ast}\rangle) ( 1 - P_I(|n_I^{\ast}\rangle) \\
	&=& \delta_{IJ} \kappa^2 n_I^{\ast} (F - 1),
\end{eqnarray}
where $\delta_{IJ}$ is a Kronecker delta, and the relation $F P_I(|n_I^{\ast}\rangle) = 1$ is used to derive the second equation. 
The relation between $\mathbf{G}$, $\mathbf{S}$, and $\mathbf{H}$ is 
\begin{equation}
	\mathbf{G} - \mathbf{S}\mathbf{G}\mathbf{S}^{\rm T} = \mathbf{H}, 
\end{equation}
where the superscript ${\rm T}$ denotes the transpose of the matrix. 
Although $\mathbf{G}$ is not simply expressed by the known matrices $\mathbf{S}$ and $\mathbf{H}$, 
it is written in a series as 
\begin{equation}
	\mathbf{G} = \mathbf{H} + \mathbf{SHS}^{\rm T} + \mathbf{SSHS}^{\rm T}\mathbf{S}^{\rm T} + \cdots. 
\end{equation}
Hence, $\mathbf{G}$ is calculated by the following iterations: 
\begin{equation}
	\mathbf{G}_k = \mathbf{H} + \mathbf{S}\mathbf{G}_{k-1}\mathbf{S}^{\rm T}, 
\end{equation}
where $\mathbf{G}_0 = \mathbf{H}$. 
We repeated this iteration until the absolute values of all the elements of the matrix $(\mathbf{G}_k - \mathbf{G}_{k-1})$ are less than $10^{-3}$.
The number of iterations is typically of order $10^4$. (It depends on the species composition.)
Since the eigenvalues of $\mathbf{S}$ are close to unity, 
many iterations are necessary to obtain an accurate estimate of $\mathbf{G}$. 

\section{Linear stability analysis for Model A}\label{sec:a_stability}
\begin{figure}[ht!]
\begin{center}
\subfigure{
\includegraphics[width=8.0cm]{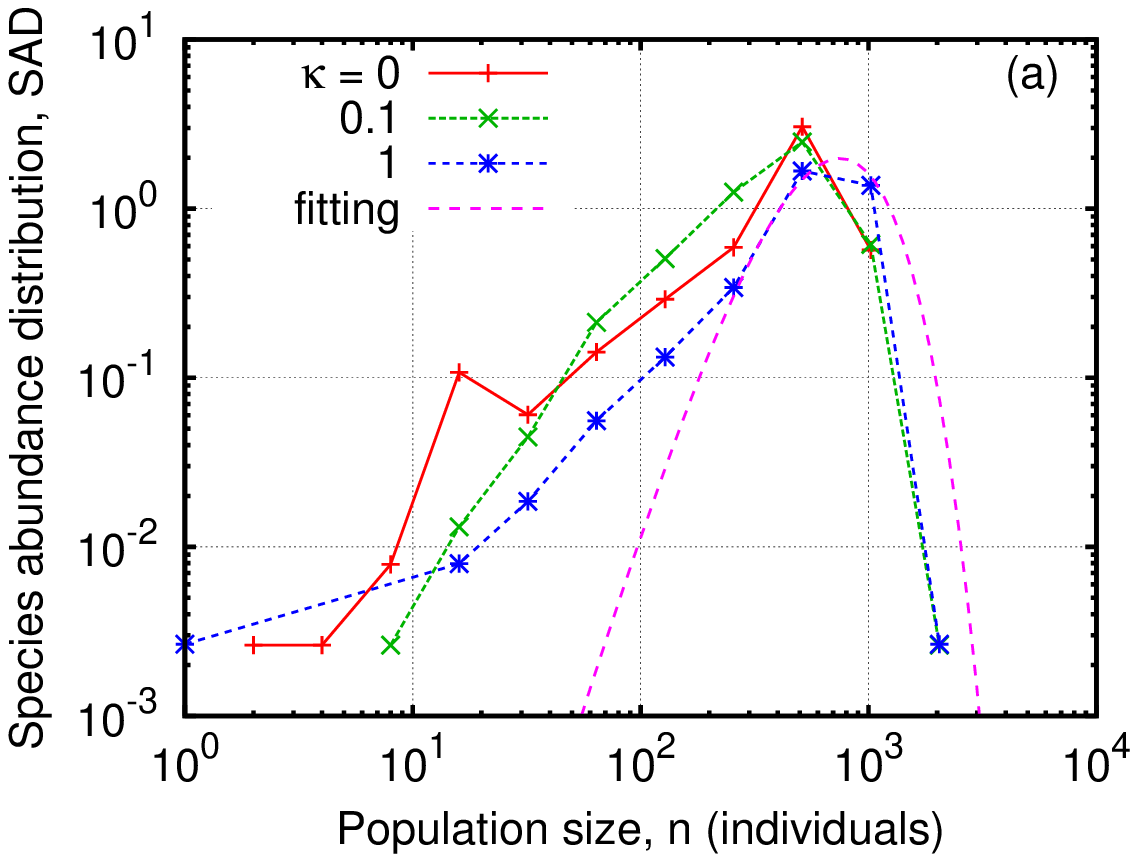}
\label{fig:a-sad}
}
\subfigure{
\includegraphics[width=8.0cm]{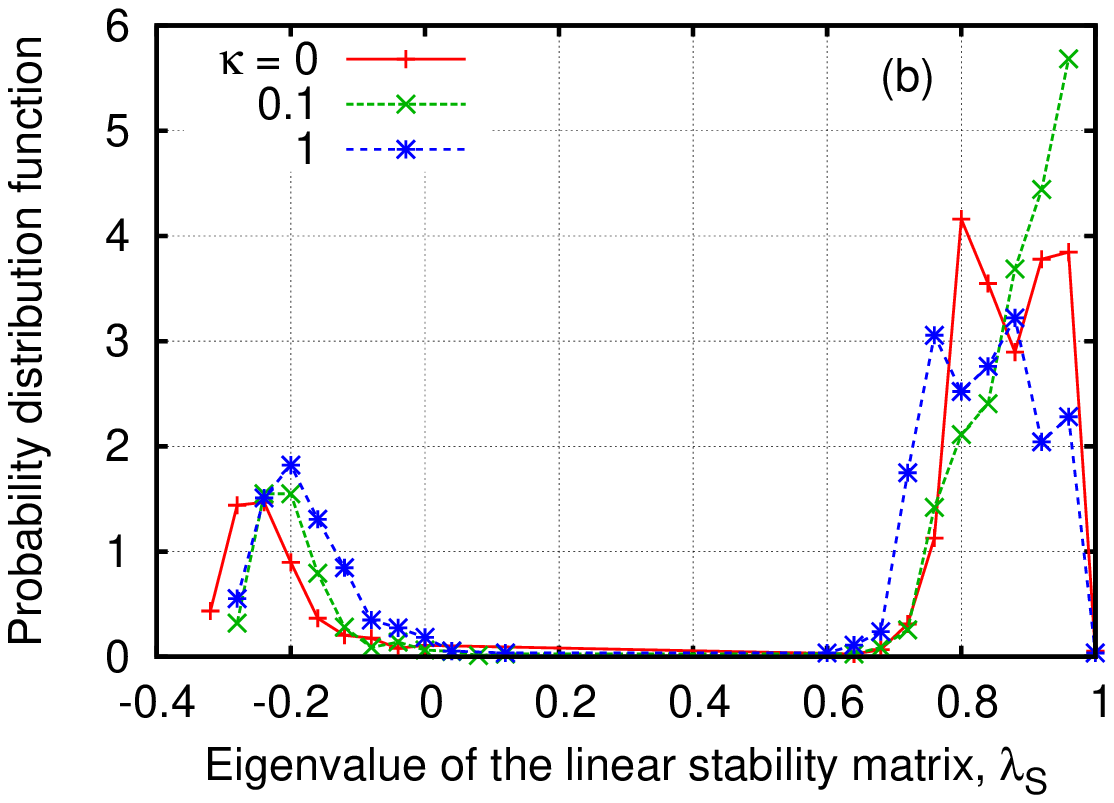}
\label{fig:a-s-eigens}
}
\subfigure{
\includegraphics[width=8.0cm]{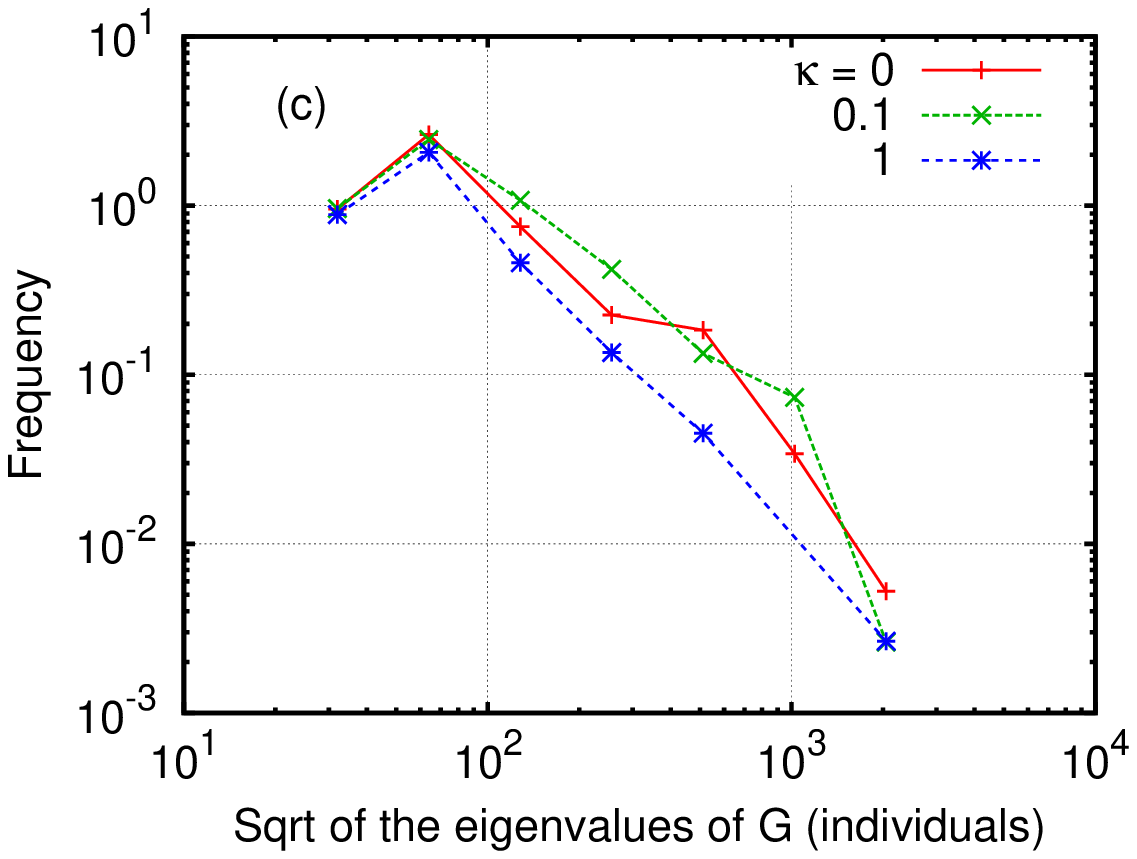}
\label{fig:a-g-eigens}
}
\end{center}
\caption{
(Color online) 
(a) Species abundance distribution (SAD), 
(b) probability distribution functions (pdf) of the eigenvalues of the linear stability matrix, $\lambda_{\mathbf{S}}$,  
(c) pdf of the eigenvalues of the covariance matrix, $\lambda_{\mathbf{G}}$,  
 for Model A with several $\kappa$. The data are sampled every one million generations. 
The SADs show the number of species whose populations are within each bin region, for each community. 
The fitting function for the SAD ($\kappa=1$) is Eq.~(\ref{eq:sad}) with $\beta = 4.5$ and $\gamma = 0.006$.
Compare with Fig.~\ref{fig:b-fluctuations} for Model B.
\label{fig:a-sads} }
\end{figure}

SAD and the linear stability matrix $\mathbf{S}$ for Model A were calculated. 
Figure \ref{fig:a-sad} shows the SAD for Model A at several noise levels.
The data are sampled every one million generations after an initial warm-up period of four million generations. 
In the same way as for Model B, we first removed unsuccessful mutants 
and obtained the core fixed-point communities. 
The profiles show little dependence on the noise level. 
We tried fitting the SADs by Eq.~(\ref{eq:sad}), but the fitting does not look very reasonable. 
The eigenvalue distribution of the linear stability matrix $\mathbf{S}$ 
and the distribution of square roots of the eigenvalues of the covariance matrix $\mathbf{G}$ are shown in 
Fig.~\ref{fig:a-s-eigens} and Fig.~\ref{fig:a-g-eigens}, respectively. 
Figure \ref{fig:a-s-eigens} shows that the linear stability for Model A is much stronger than for Model B. 
Negative eigenvalues of $S$ correspond to oscillating modes. 
The amplitude of fluctuations that would appear under the noise were estimated and 
shown in Fig.~\ref{fig:a-g-eigens}. 
The typical size of the amplitude is smaller than the typical population sizes 
and, as a consequence, the resident species seldom suffer from the noise effects.

\section{Trial for Model B with constant birth cost}\label{sec:b-b-const}
\begin{figure}[ht!]
\begin{center}
\subfigure{
\includegraphics[width=8.0cm]{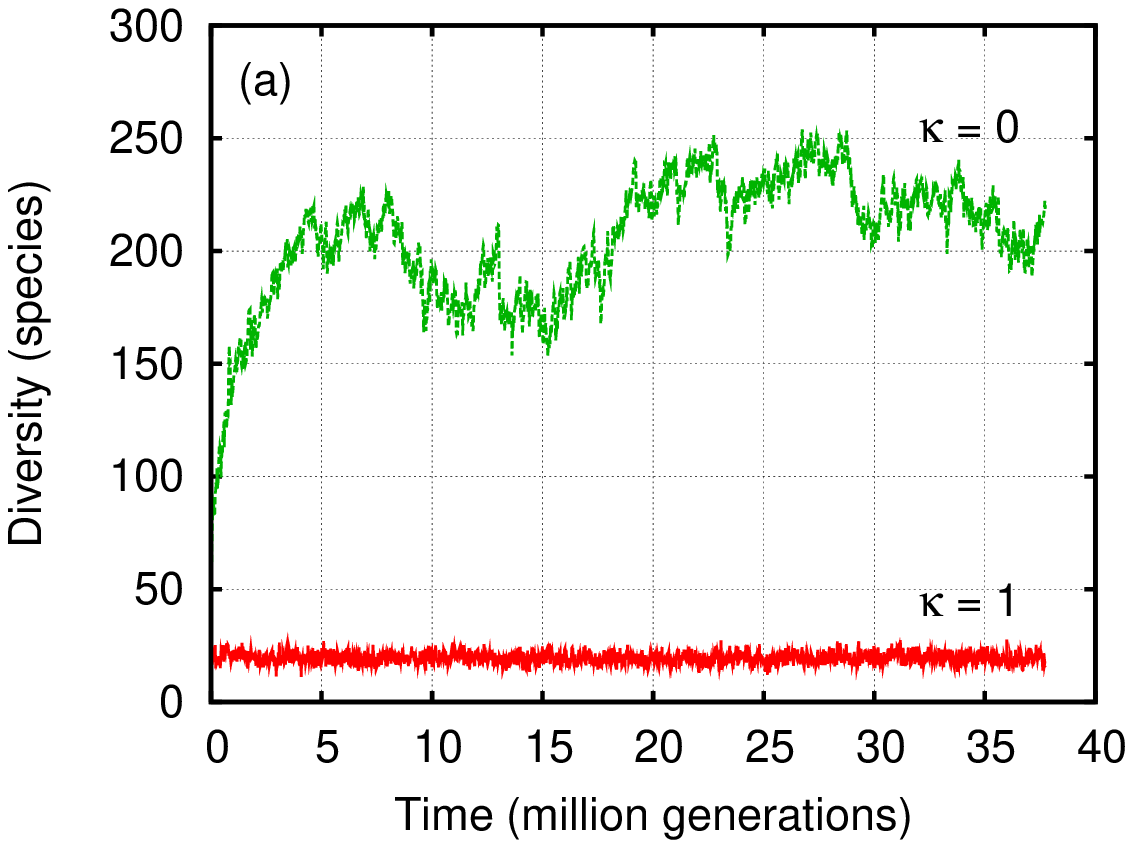}
\label{fig:cB-div-time}
}
\subfigure{
\includegraphics[width=8.0cm]{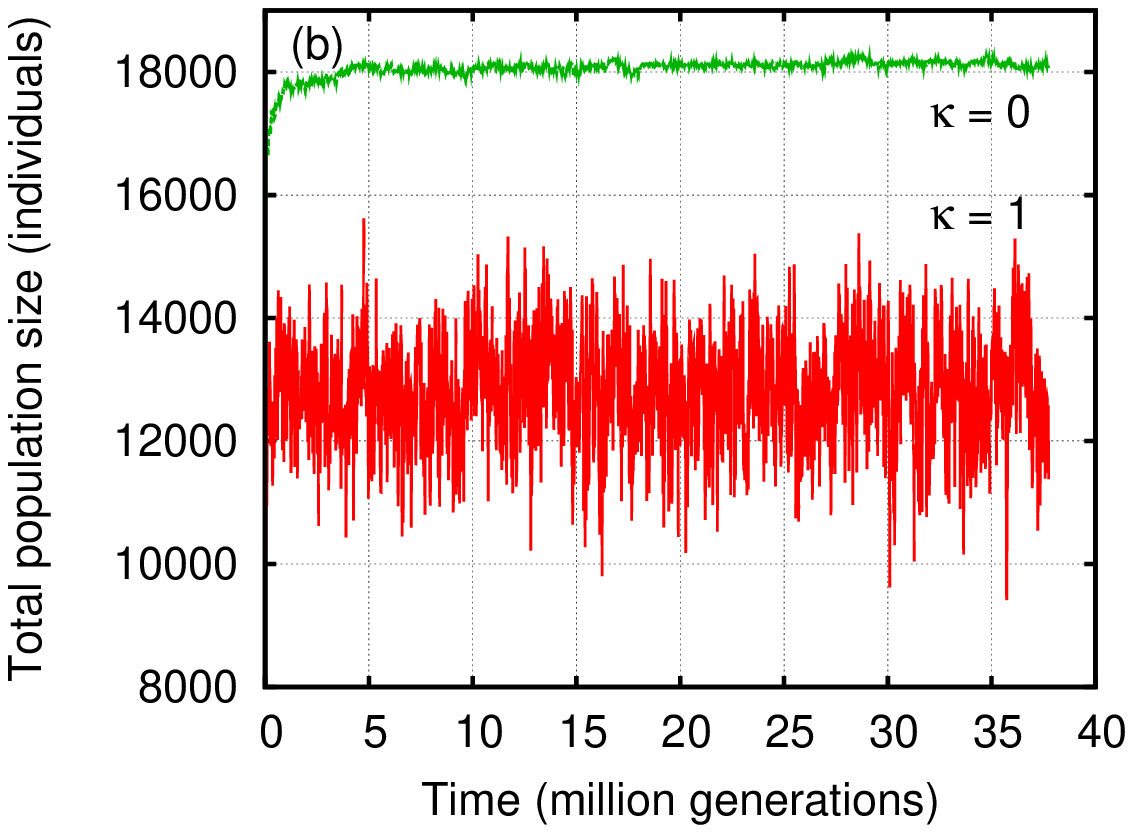}
\label{fig:cB-biomass-time}
}
\end{center}
\caption{
(Color online) Time series of (a) exponential Shannon-Wiener diversity and (b) total population size 
for the modified Model B with $\mu = 0.001$ and $L=20$ at $\kappa = 0$ and $1$. 
The data are plotted every $16384$ generations.
Compare to Fig.~\ref{fig:b-mut-div-biomass}.
\label{fig:cB-time-series}}
\end{figure}

The distribution of the birth cost $b_I$ for Model B at $\kappa = 0$ has a quite sharp peak close to zero. 
This emerges as a result of the selection and causes low linear stability. 
Since quite low birth cost is not very realistic, 
we also tried a model in which the birth cost is fixed to be $0.1$ for all species. 
Typical time series of diversity and total population size for $\mu = 0.001$ and $L = 20$ are shown in Fig.~\ref{fig:cB-time-series}. 
Averages of the diversity index $D$ for $\kappa = 0$ and $1$ are $219$ and $19.7$, respectively. 
Thus, there are significant decreases in diversity and total population size with increasing noise. 

\begin{figure}[ht!]
\begin{center}
\subfigure{
\includegraphics[width=8.0cm]{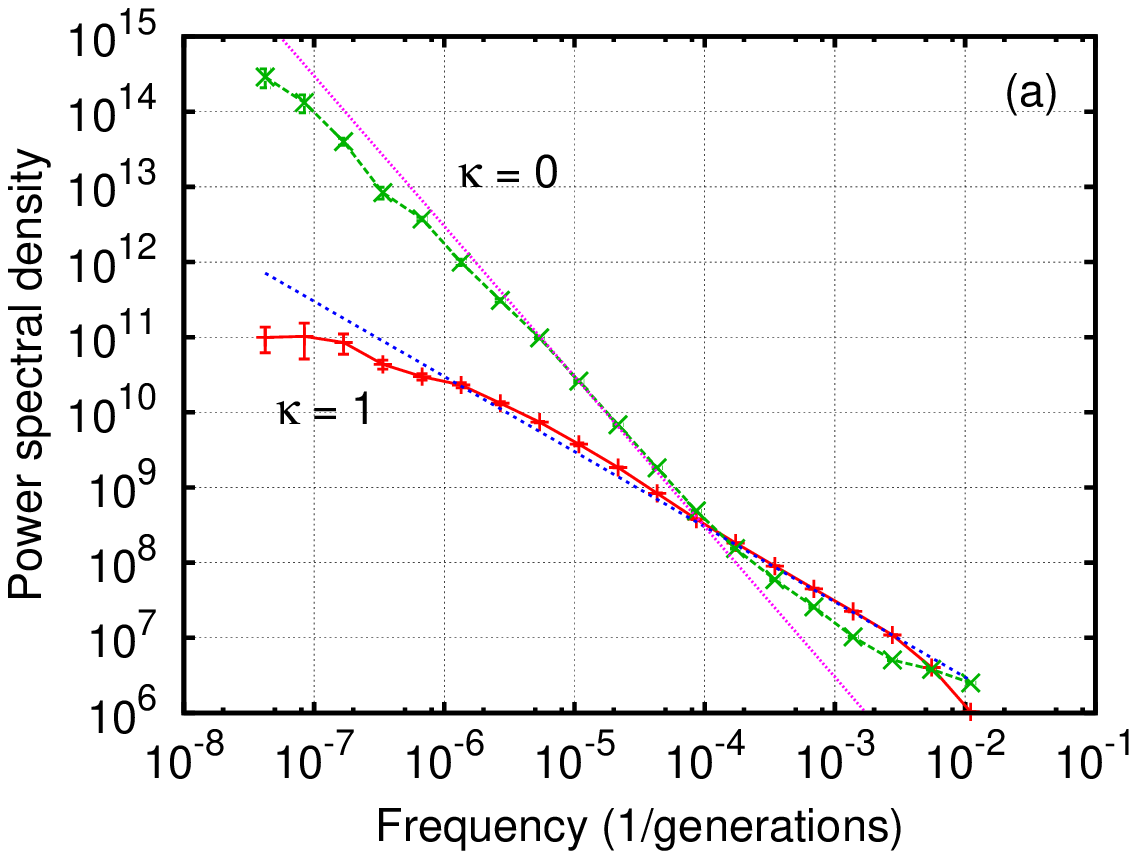}
\label{fig:cB-psd-div}
}
\subfigure{
\includegraphics[width=8.0cm]{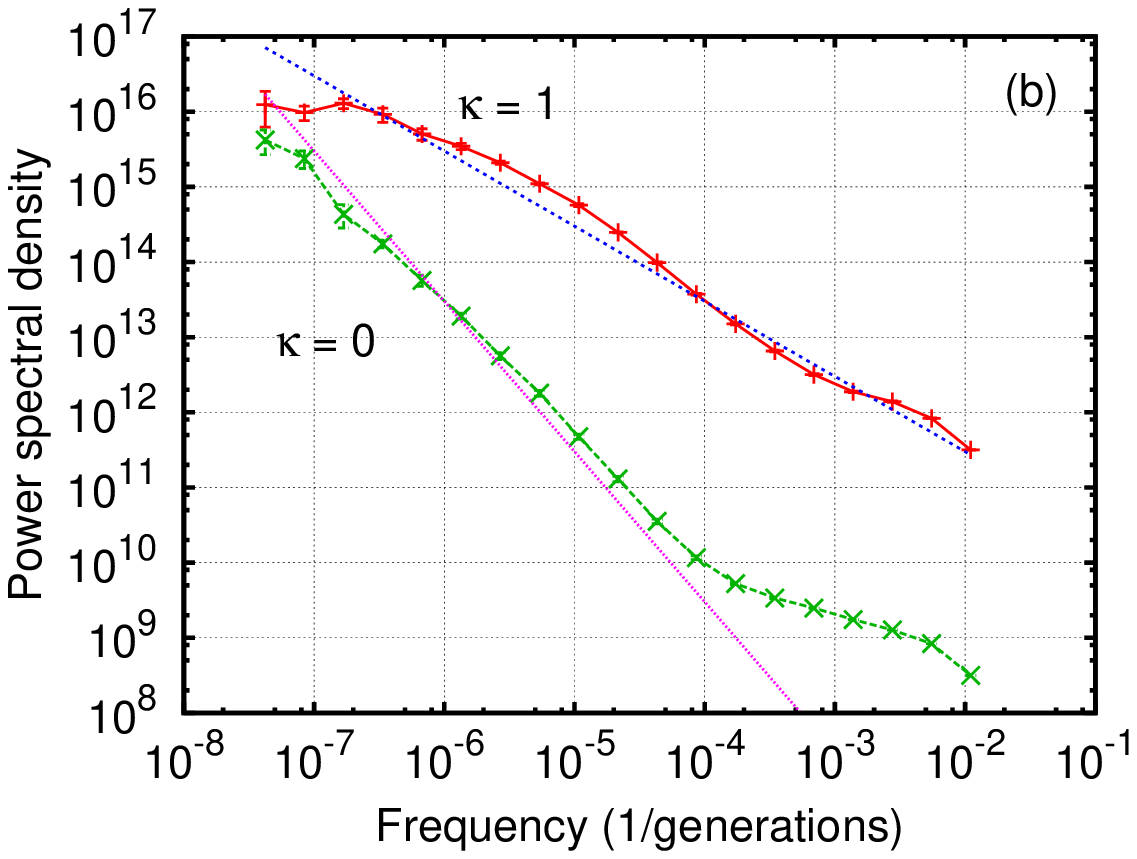}
\label{fig:cB-psd-biomass}
}
\end{center}
\caption{
(Color online) PSDs of (a) exponential Shannon-Wiener diversities and (b) total population sizes 
for the modified Model B. 
In both figures, lines corresponding to $1/f$ and $1/f^2$ are shown as guides to the eye.
Data are averaged over six independent runs, and their statistical errors are also shown.
Compare to Fig.~\ref{fig:b-mut-psds}.
\label{fig:cB-psds}}
\end{figure}

\begin{figure}[ht!]
\begin{center}
\subfigure{
\includegraphics[width=8.0cm]{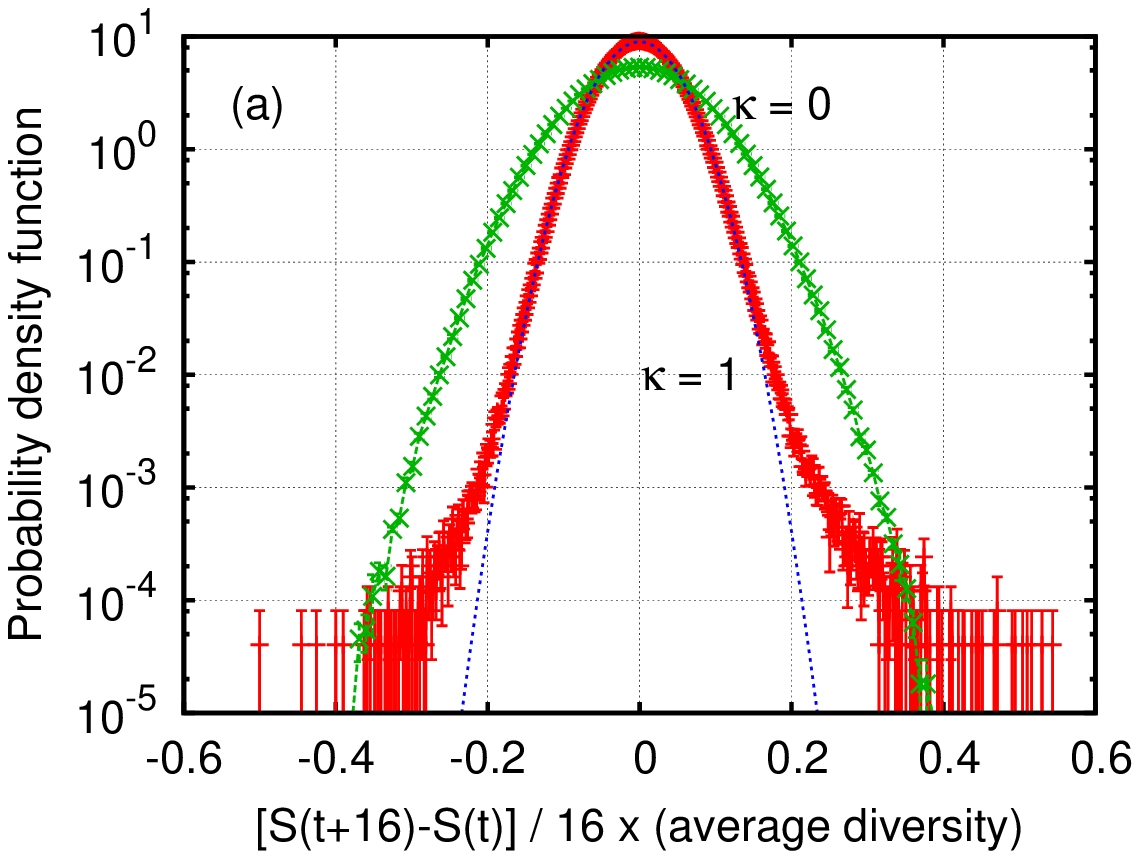}
\label{fig:cB-pdf-dsdt}
}
\subfigure{
\includegraphics[width=8.0cm]{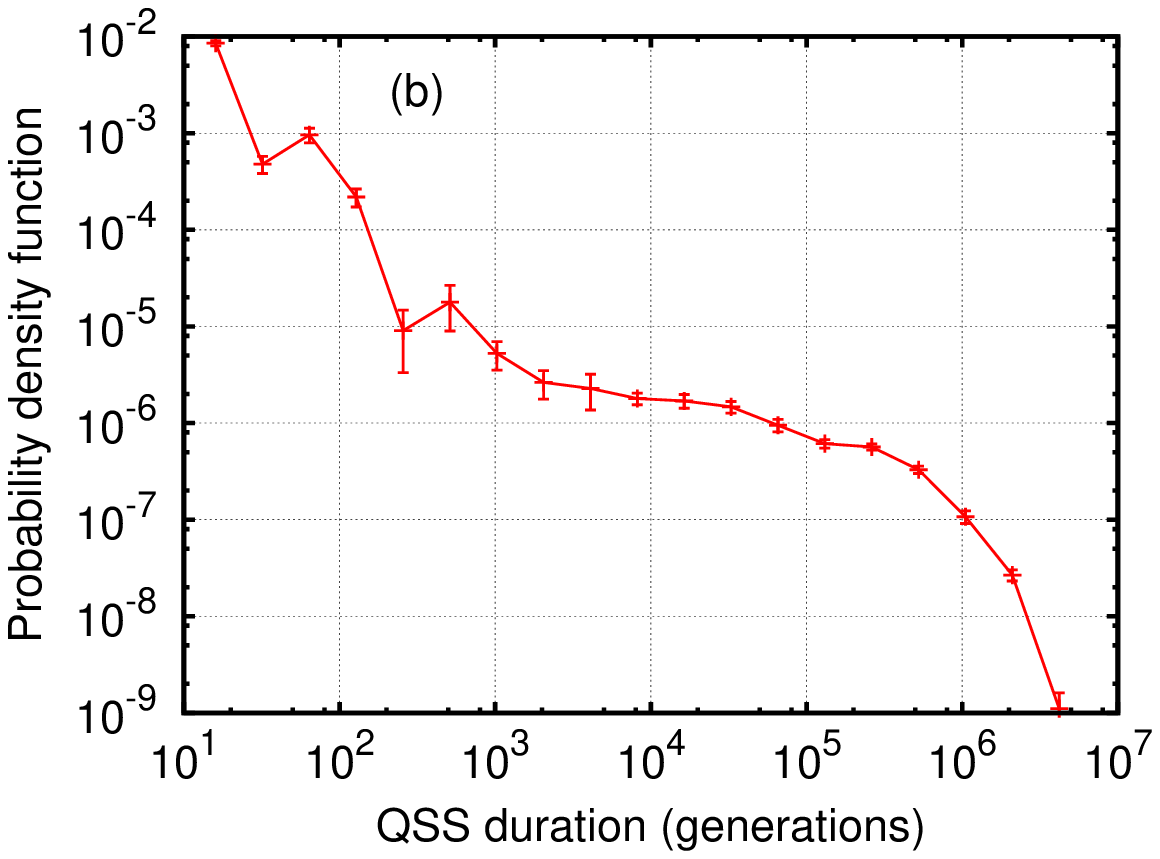}
\label{fig:cB-pdf-duration}
}
\subfigure{
\includegraphics[width=8.0cm]{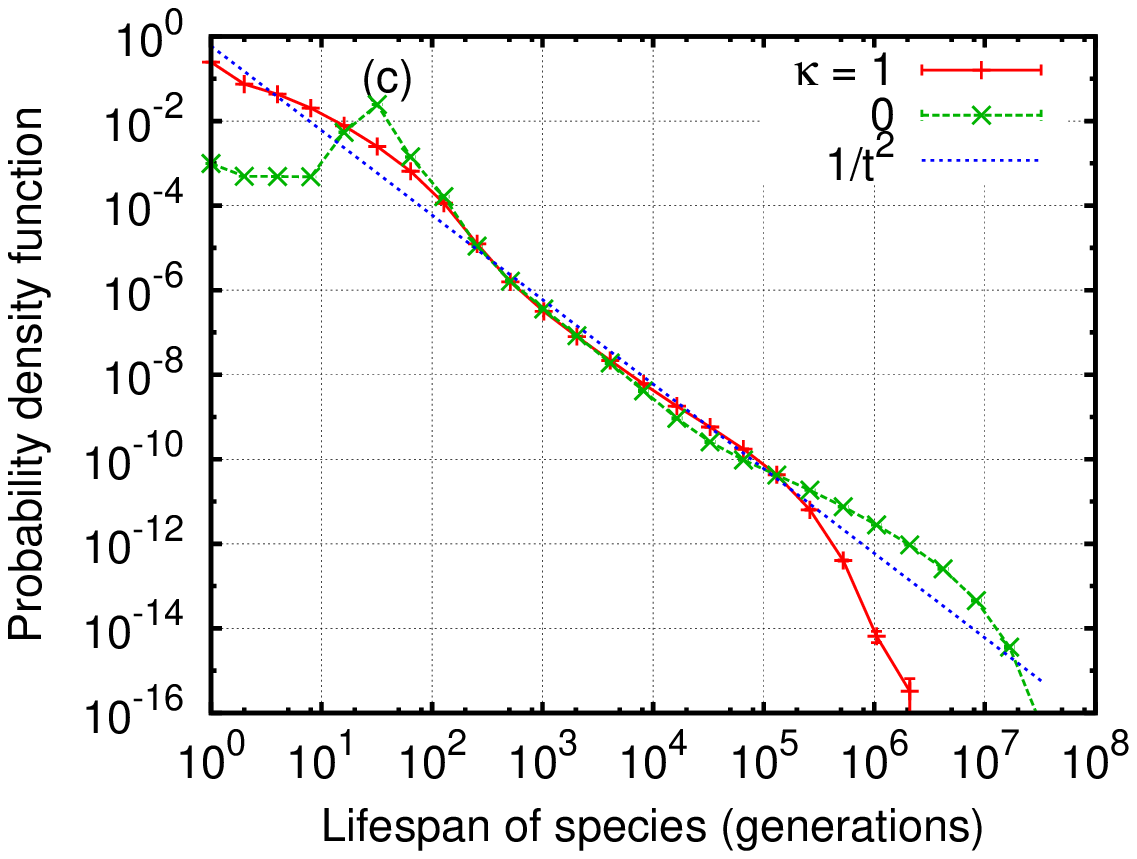}
\label{fig:cB-lifespan}
}
\end{center}
\caption{
(Color online) 
(a) Probability density function of $[S(t+16)-S(t)]/16\times \bar{D}$, where $\bar{D}$ is the average diversity, 
for the modified Model B with $\kappa = 0$ and $1$. 
A Gaussian function is also shown as a guide to the eye.
Compare to Fig.~\ref{fig:b-mut-dsdt-normed}.
(b) The QSS duration distribution for $\kappa = 1$. 
The cutoff to detect QSS is $0.012$.
Compare to Fig.~\ref{fig:b-mut-duration}.
(c) Species lifetime distribution for $\kappa = 0$ and $1$. 
A power law $t^{-2}$ is also shown as a guide to the eye.
Compare to Fig.~\ref{fig:b-mut-lspan}.
\label{fig:cB-durations}}
\end{figure}

\begin{figure}[ht!]
\begin{center}
\subfigure{
\includegraphics[width=8.0cm]{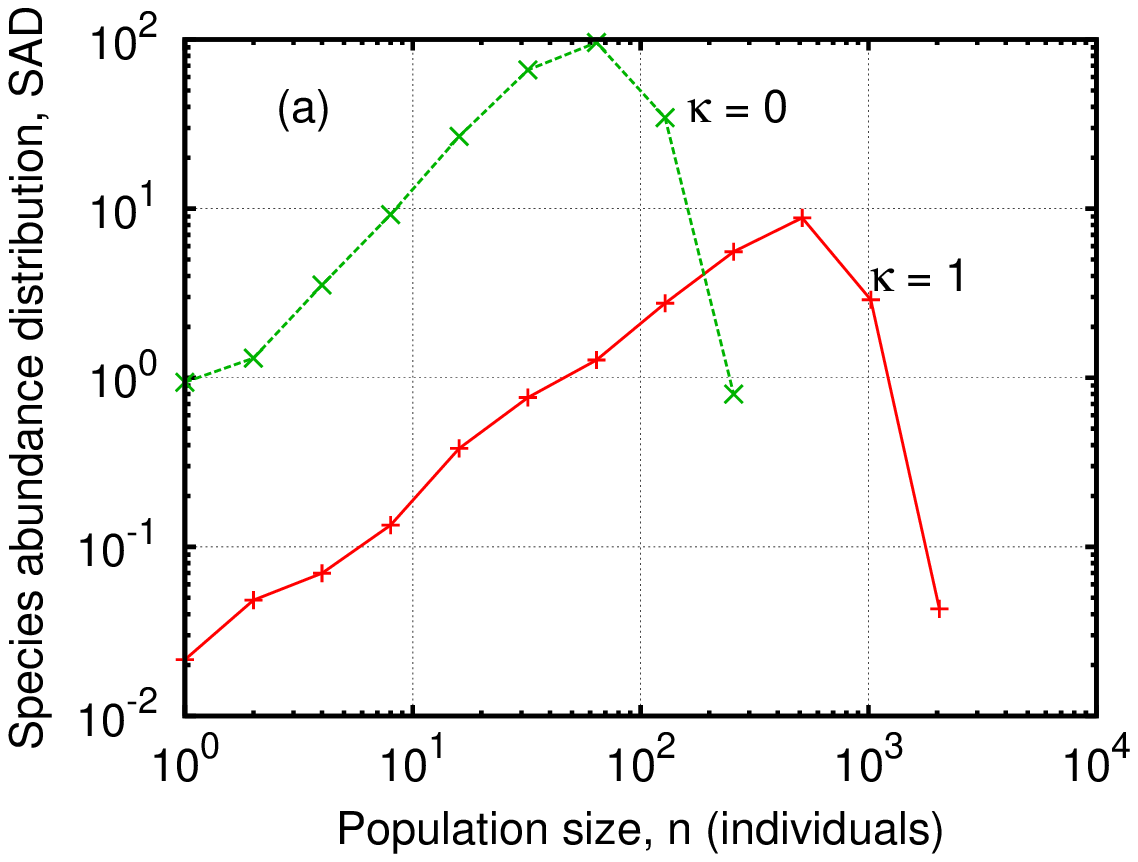}
\label{fig:cB-SAD}
}
\subfigure{
\includegraphics[width=8.0cm]{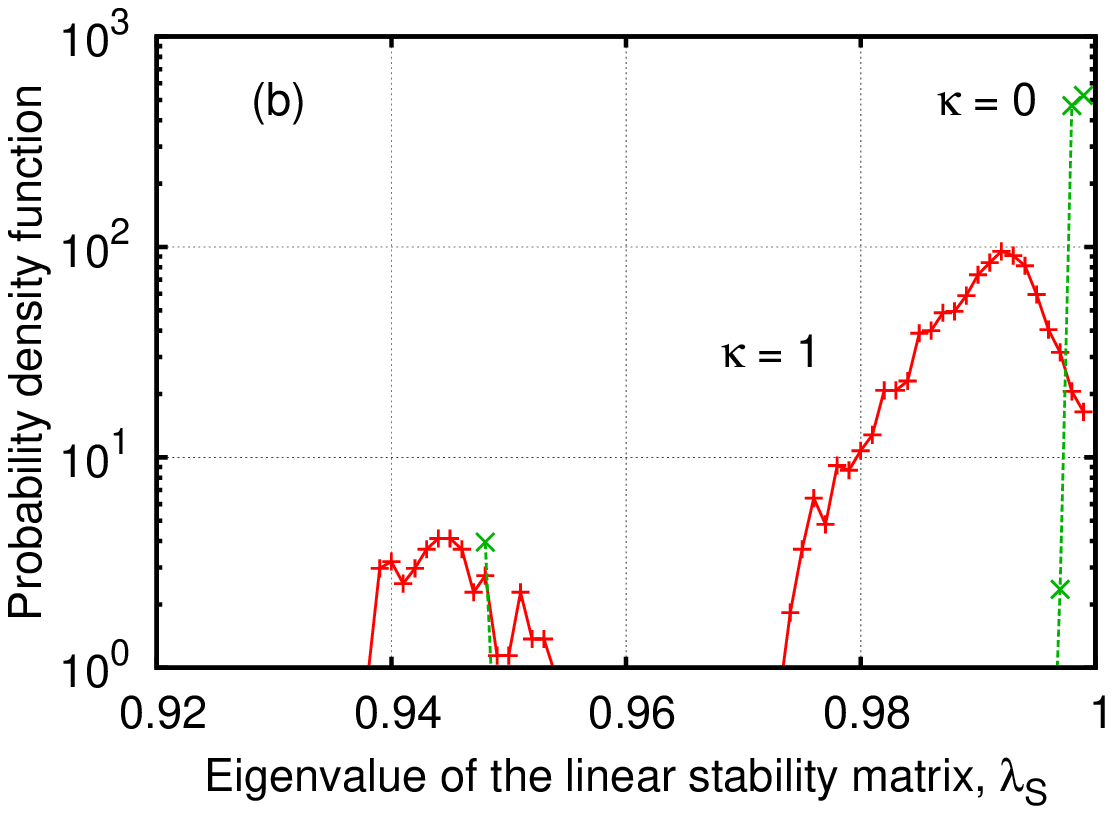}
\label{fig:cB-s-eigen}
}
\subfigure{
\includegraphics[width=8.0cm]{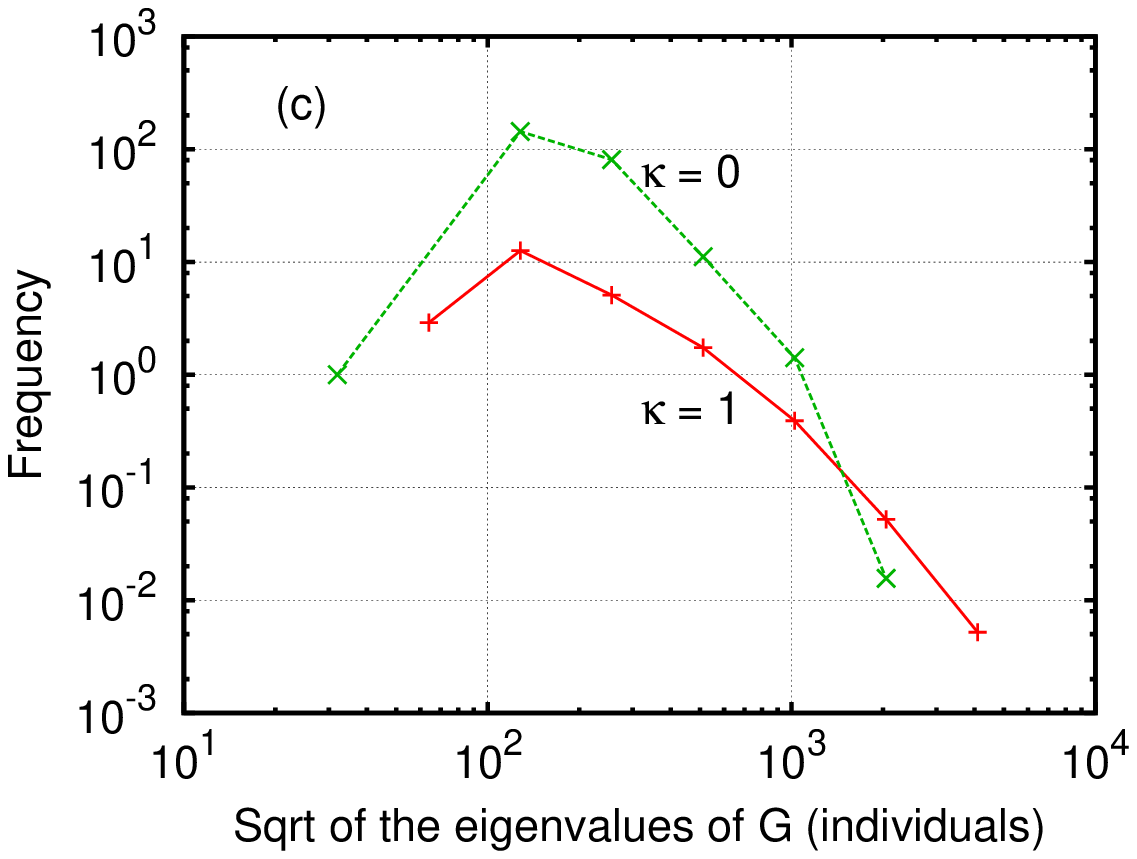}
\label{fig:cB-g-eigen}
}
\end{center}
\caption{
(Color online) 
(a) Species abundance distribution for the modified Model B with $\kappa = 0$ and $1$.
(b) The distribution of the eigenvalues of the linear stability matrix, $\lambda_{\mathbf{S}}$ for the modified Model B with $\kappa = 1$ and $0$.
(c) The distribution of the square roots of the eigenvalues of the covariance matrix, $\lambda_{\mathbf{G}}$.
The data are shown in the same way as the SADs, (binning in $\log_2$ scale).
Compare to Fig.~\ref{fig:b-fluctuations}.
\label{fig:cB-SADs}}
\end{figure}

\begin{figure}[ht!]
\begin{center}
\subfigure{
\includegraphics[width=8.0cm]{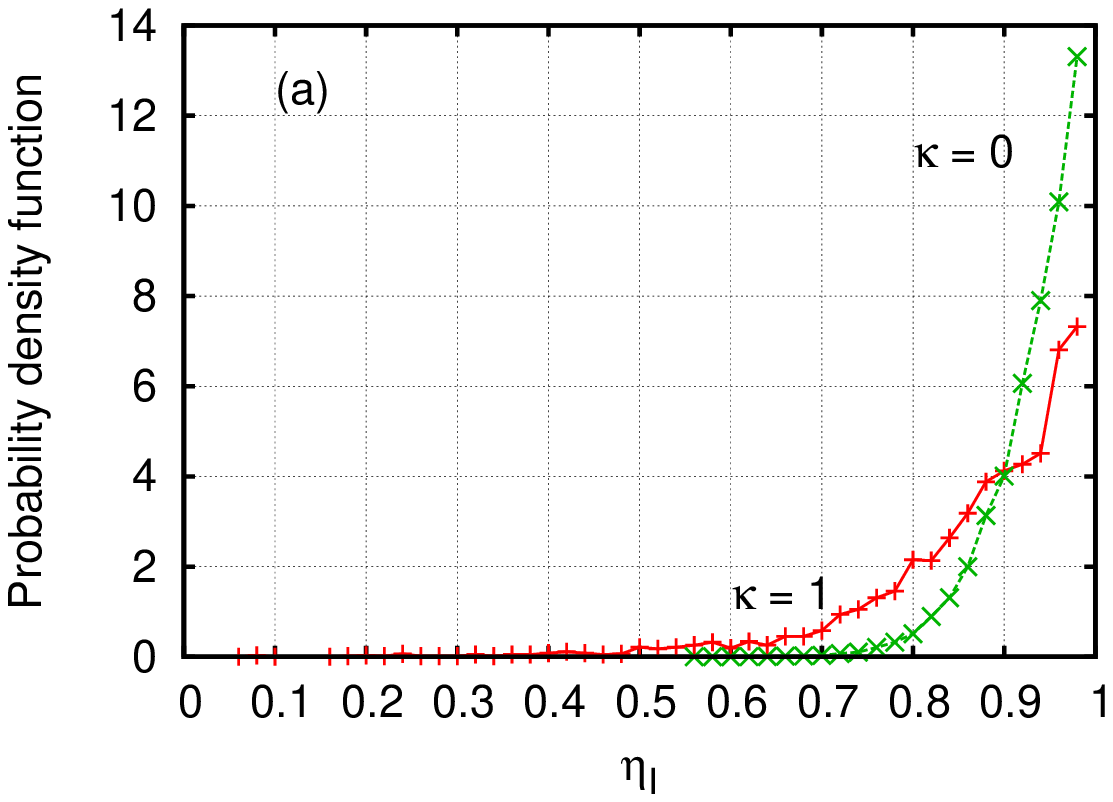}
\label{fig:cB-eta-histo}
}
\subfigure{
\includegraphics[width=8.0cm]{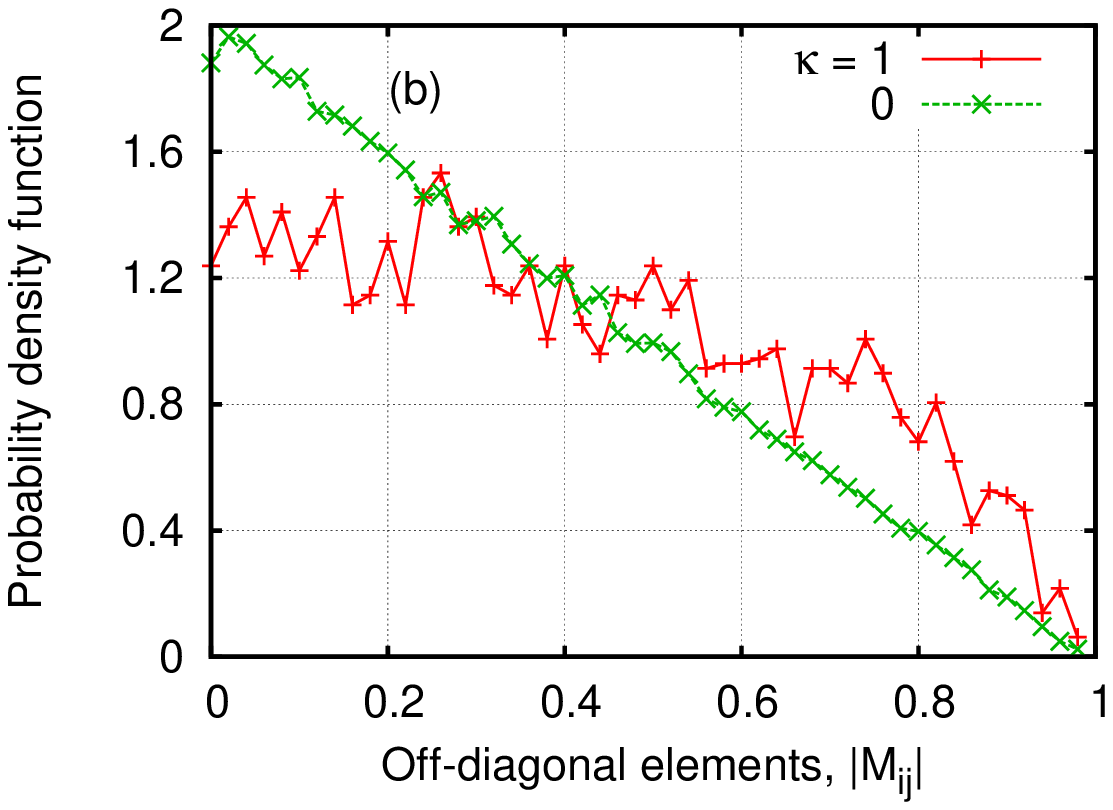}
\label{fig:cB-mij-histo}
}
\end{center}
\caption{
(Color online) 
(a) The distribution of $\eta_{I}$. Species with large $\eta_I$ are more favored in the limit without demographic noise.
(b) The distribution of the off-diagonal elements of the interaction matrix, $|M_{IJ}|$.
Compare to Fig.~\ref{fig:b-elements-pdf}.
\label{fig:cB-elements-pdf}}
\end{figure}

Properties of long-term fluctuations are also estimated. 
Six independent simulations of $2^{25}$ generations with $2^{22}$ warm-up generations were performed.
Figure \ref{fig:cB-psds} shows the PSDs of diversity and total population sizes. 
For $\kappa = 0$ and $1$, approximate $1/f^2$ and $1/f$ fluctuations are observed respectively. 
This dependence on the noise level is similar to the original Model B. 
In the same way as the original models, the distribution of the logarithmic derivative of the diversity and 
the duration distribution of quiet periods are calculated. 
Under the noise, the distribution of $dS/dt$ has a Gaussian center and wider wings [Fig.~\ref{fig:cB-pdf-dsdt}].
The duration of quiet periods distributes broadly although it is not a power-law distribution [Fig.~\ref{fig:cB-pdf-duration}]. 
This is characteristic of the modified model. 
The species-lifetime distribution shows approximate $1/t^2$ distributions regardless of the noise levels [Fig.~\ref{fig:cB-lifespan}]. 
The SADs [Fig.~\ref{fig:cB-SAD}], the eigenvalue distributions of $S$ and $G$ [Fig.~\ref{fig:cB-s-eigen} and \ref{fig:cB-g-eigen}], 
and the distribution of $M_{IJ}$ [Fig.~\ref{fig:cB-mij-histo}] show 
the same dependence on $\kappa$ as the original Model B.
The distribution of $\eta_{I}$ has a different dependence on the noise level than the original Model B [Fig.~\ref{fig:cB-eta-histo}]. 
When the noise is not applied, much stronger evolution pressure is applied to $\eta_{I}$.
Although there are several deviations from the results for the original Model B, 
qualitatively the same behaviors are observed.

\end{document}